\documentclass[12pt]{article}
\usepackage[dvips]{epsfig}
\usepackage{epsf}
\newcommand{\beq}{\begin{equation}}
\newcommand{\eps}{\epsilon}
\newcommand{\eeq}{\end{equation}}
\newcommand{\bea}{\begin{eqnarray}}
\newcommand{\eea}{\end{eqnarray}}
\begin{document}

\begin{center}

{\large \bf 
MERGING THE  CEM2K AND LAQGSM CODES WITH GEM2 TO DESCRIBE FISSION AND
LIGHT-FRAGMENT PRODUCTION}\\

\vspace{0.8cm}

{\bf S. G. Mashnik$^1$, K. K. Gudima$^{2}$, and A. J. Sierk$^1$}\\ 

\vspace{0.8cm}
$^1$Los Alamos National Laboratory,  Los Alamos, NM 87545, USA\\

\vspace{0.8cm}
$^2$ Institute of Applied Physics,
Academy of Science of Moldova, Kishinev, MD-2028, Moldova\\

\vspace{0.8cm}
{\bf Abstract}
\end{center}
We present the current status of the improved 
Cascade-Exciton Model (CEM) code CEM2k and of 
the Los Alamos version of the Quark-Gluon String Model code LAQGSM.
To describe fission and light-fragment (heavier than He4) production, 
both CEM2k and LAQGSM have been merged with the GEM2 code of 
Furihata. We present some results on proton- and deuteron-induced
spallation, fission, and fragmentation reactions predicted by these 
extended versions of CEM2k and LAQGSM.
We show that merging CEM2k and LAQGSM with GEM2 allows us to
describe many fission and fragmentation reactions in addition to the 
spallation reactions which are already relatively well described.
Nevertheless, the standard version of GEM2 does not
provide a completely satisfactory description of complex particle spectra, 
heavy-fragment emission, and spallation yields, and is not yet a reliable 
tool for applications. 
We conclude that we may choose to use a model similar to the GEM2 approach 
in our codes,
but it must be significantly extended and further improved.
We observe that it is not sufficient to analyze only A and Z distributions 
of the product yields when evaluating this type of model, 
as is often done in the literature; instead it is important 
to study all the separate isotopic yields as well as the spectra of 
light particles and fragments.

\newpage

{\noindent \bf Introduction}\\

During recent years, for a number of applications like 
Accelerator Transmutation of nuclear Wastes (ATW),
Accelerator Production of Tritium (APT),
Rare Isotope Accelerator (RIA), Proton Radiography (PRAD),
and others projects,
we have developed
at the Los Alamos National Laboratory
an improved version of the 
Cascade-Exciton Model (CEM), 
contained 
in the code CEM2k, to describe 
nucleon-induced reactions at incident energies up to 5 GeV 
\cite{CEM2k}
and the Los Alamos version of the Quark-Gluon String Model,
realized
in the high-energy code LAQGSM \cite{LAQGSM},
able to describe both particle- and nucleus-induced reactions
at energies up to about 1 TeV/nucleon.

In our original motivation, different versions of the CEM and
LAQGSM codes were 
developed to reliably describe the yields of spallation products
and spectra of secondary particles, without a special emphasis
on 
complex-particle and light-fragment emission or on
fission fragments in reactions with heavy targets. 
In fact, the initial versions of the CEM2k and LAQGSM codes
simulate spallation 
only and do not calculate the process of fission, and do
not provide fission fragments and a further possible evaporation of
particles from them. Thus, in simulating the compound stage of a 
reaction, when these codes encounter a fission, they simply tabulate 
this event (that permits calculation of fission cross sections and 
fissility) and finish the calculation of this event without a subsequent 
treatment of fission fragments.  To be able to describe 
nuclide production in the fission region, these codes have to be extended
by incorporating a model of high energy fission
({\it e.g.}, in the transport code MCNPX \cite{MCNPX}, 
where CEM2k and, initially,
its precursor, CEM97 \cite{CEM97}, are used, 
they are supplemented by Atchison's fission model \cite{RAL,RAL1}.

Since many nuclear and astrophysical applications require reliable
data also on complex particles (gas production) and light and/or fission
fragment production, we addressed these questions 
by further development of CEM2k and LAQGSM codes.
We tried different ways of solving these problems and 
as a first attempt to describe with our codes both emission of 
intermediate-mass fragments heavier than $^4$He and production of 
heavy fragments from fission, we merged CEM2k and LAQGSM
with the Generalized Evaporation Model
(GEM) code by Furihata \cite{GEM2,GEM2a}. 
We have benchmarked our codes on all 
proton-nucleus and nucleus-nucleus 
reactions measured recently at 
GSI (Darmstadt, Germany) 
and on many other different
reactions at lower and higher energies measured 
earlier at other laboratories.
The size of the present paper allows us to present here only
a few results, which we choose to be
for the GSI measurements on
interaction of $^{208}$Pb beams with p \cite{Enqvist01}
and d \cite{Enqvist02} targets.
Results for other reactions may be found in 
\cite{SantaFe02,COSPAR02}.\\

{\noindent \bf CEM2k and LAQGSM Codes}\\

A detailed description of the initial version of the CEM may be found
in Ref. \cite{CEM}, therefore we outline here only its basic
assumptions.
The CEM assumes that reactions occur in three stages. The first
stage is the IntraNuclear Cascade (INC) 
in which primary particles can be re-scattered and produce secondary
particles several times prior to absorption by or escape from the nucleus.
The excited residual nucleus remaining after the 
cascade determines the particle-hole configuration that is
the starting point for the preequilibrium stage of the
reaction. The subsequent relaxation of the nuclear excitation is
treated in terms of an improved Modified Exciton Model (MEM) of 
preequilibrium 
decay followed by the equilibrium evaporative final stage of the reaction.
Generally, all three stages contribute to experimentally measured outcomes.

The improved cascade-exciton model in the code CEM2k differs from 
the older
CEM95 version 
(which is available free from the NEA/OECD, Paris)
\cite{CEM95}
by incorporating new 
approximations for the elementary cross sections used in the cascade,
using more precise values for nuclear masses and 
pairing energies, 
employing a
corrected systematics for the level-density
parameters, 
adjusting the cross sections for pion absorption on quasi-deuteron 
pairs inside a nucleus, 
including the Pauli principle 
in the preequilibrium calculation, 
and improving the calculation of the fission widths.
Implementation of significant refinements 
and improvements in the algorithms of many subroutines 
led to a decrease of
the computing time by up to a
factor of 6 for heavy nuclei, which 
is very important when performing
simulations with transport codes.
Essentially, CEM2k \cite{CEM2k} has a longer cascade stage,
less preequilibrium emission, and a longer evaporation stage
with a higher excitation energy, as compared to its precursors
CEM97 \cite{CEM97} and CEM95 \cite{CEM95}.
Besides the changes to CEM97 and CEM95 mentioned above, we also made a 
number of other improvements and refinements, such as:
(i)
imposing momentum-energy conservation for each simulated event
(the Monte Carlo algorithm previously used in CEM 
provides momentum-energy conservation only 
statistically, on the average, but not exactly for the cascade stage 
of each event),
(ii)
using real binding energies for nucleons at the cascade 
stage instead of the approximation of a constant
separation energy of 7 MeV used in previous versions of the CEM,
(iii)
using reduced masses of particles in the calculation of their
emission widths instead of using the approximation
of no recoil used previously, and
(iv)
a better approximation of the total reaction cross sections.
On the whole, this set of improvements leads to a much better description
of particle spectra and yields of residual nuclei and a better 
agreement with available data for a variety of reactions.
Details, examples, and further references may be found in Refs.
\cite{CEM2k,Titarenko02}. 

The Los Alamos version of the Quark-Gluon String Model
(LAQGSM) \cite{LAQGSM} is the next generation of 
the Quark-Gluon String Model (QGSM) by Amelin {\it et al.}
(see \cite{QGSM} and references therein) and is intended to describe
both particle- and nucleus-induced reactions at energies up to
about 1 TeV/nucleon. 
The core of the QGSM is built on a time-dependent version of the
intranuclear cascade model developed at Dubna,
often referred in the literature simply
as the Dubna intranuclear Cascade Model (DCM) (see \cite{Toneev83}
and references therein).
The DCM models interactions of fast cascade particles (``participants")
with nucleon spectators of both the target and projectile nuclei and
includes 
interactions of two participants (cascade particles)
as well.
It uses experimental cross sections (or those calculated by the Quark-Gluon 
String Model for energies above 4.5 GeV/nucleon) for these
elementary interactions to simulate angular and energy distributions
of cascade particles, also considering the Pauli exclusion
principle. When the cascade stage of a reaction is completed, QGSM uses the
coalescence model described in \cite{Toneev83}
to ``create" high-energy d, t, $^3$He, and $^4$He by
final state interactions among emitted cascade nucleons, already outside 
of the colliding nuclei.
After calculating the coalescence stage of a reaction, QGSM
moves to the description of the last slow stages of the interaction,
namely to preequilibrium decay and evaporation, with a possible competition
of fission
using the standard version of the CEM \cite{CEM}. 
But if the residual nuclei have atomic numbers 
with  $A \le 13$, QGSM uses the Fermi break-up model 
to calculate their further disintegration instead of using
the preequilibrium and evaporation models.
LAQGSM differs from QGSM by replacing the preequilibrium and
evaporation parts  of QGSM described according to the standard CEM
\cite{CEM} with the new physics from CEM2k \cite{CEM2k}
and has a number of improvements 
and refinements in the cascade and Fermi break-up models (in the
current version of LAQGSM, we use the Fermi break-up model only for
 $A \le 12$). A detailed description of LAQGSM and further
references may be found in \cite{LAQGSM}.

We have benchmarked  CEM2k and LAQGSM
against most available experimental
data and have compared our results with predictions
of other current models used by the nuclear community. 
Figure 1 shows examples of calculated
neutron spectra from the interaction of protons with $^{208}$Pb at 
0.8 and 1.5 GeV compared with experimental data \cite{Ishibashi97},
while Figure 2 gives examples of neutron spectra measured
by Nakamura's group (see \cite{Iwata01} and references therein)
from 560 MeV/nucleon Ar beams on C, Cu, and Pb targets
compared with our LAQGSM results and predictions by
QMD \cite{Aichelin91} and HIC \cite{Bertini74} from
Iwata {\it et al.} \cite{Iwata01}. 
We see that our codes describe well neutron spectra both
for proton- and nucleus-nucleus reactions and agree with the
data no worse than other models do.
Similar results are obtained for other reactions for which we found 
measured data. \\

{\noindent \bf Merging CEM2k and LAQGSM with GEM2}\\

The Generalized Evaporation Model
(GEM) \cite{GEM2,GEM2a} is an extension by Furihata
of the Dostrovsky {\it et al.} \cite{Dostrovsky} evaporation model
as implemented in LAHET \cite{LAHET}
to include up to 66 types of particles and fragments that
can be evaporated from an excited compound nucleus plus a modification
of the version of Atchison's fission model \cite{RAL,RAL1}
used in LAHET. Many 
of the parameters were adjusted for a better description of 
fission reactions when using it in conjunction with the extended 
evaporation model. We merged GEM2 (the last update of the GEM code)
with CEM2k and LAQGSM as follows: we calculate the cascade and preequilibrium
stages of a reaction with our CEM2k or LAQGSM, then we describe the 
subsequent 
evaporation of particles and fragments and fission from the remaining 
excited compound nuclei using GEM2. To understand the role of 
preequilibrium particle emission, we performed calculations of all the 
reactions we tested both with emission of preequilibrium particles and 
without them, {\it i.e.}, going directly to GEM2 after the intranuclear 
cascade stage of a reaction described by CEM2k or LAQGSM.

A very detailed description of the GEM, together with a large amount
of results obtained for many reactions
using GEM coupled either with the Bertini or ISABEL INC models
in LAHET  are published by Furihata \cite{GEM2,GEM2a,GEM2mod}; many 
useful details of GEM2 may be found also in our paper \cite{SantaFe02}.
Therefore, we list briefly below
only the main features of GEM and forward readers interested in 
details to Refs. \cite{GEM2,GEM2a,SantaFe02,GEM2mod}.
Furihata did not change in GEM the general algorithms used in LAHET
to simulate evaporation and fission. The decay widths of evaporated 
particles and fragments are estimated using the classical Weisskopf-Ewing 
statistical theory \cite{ Weisskopf}.

%\newpage
\begin{figure}
\vspace{-6.0cm}
\centerline{\hspace{-2.5cm} \epsfxsize 21cm \epsffile{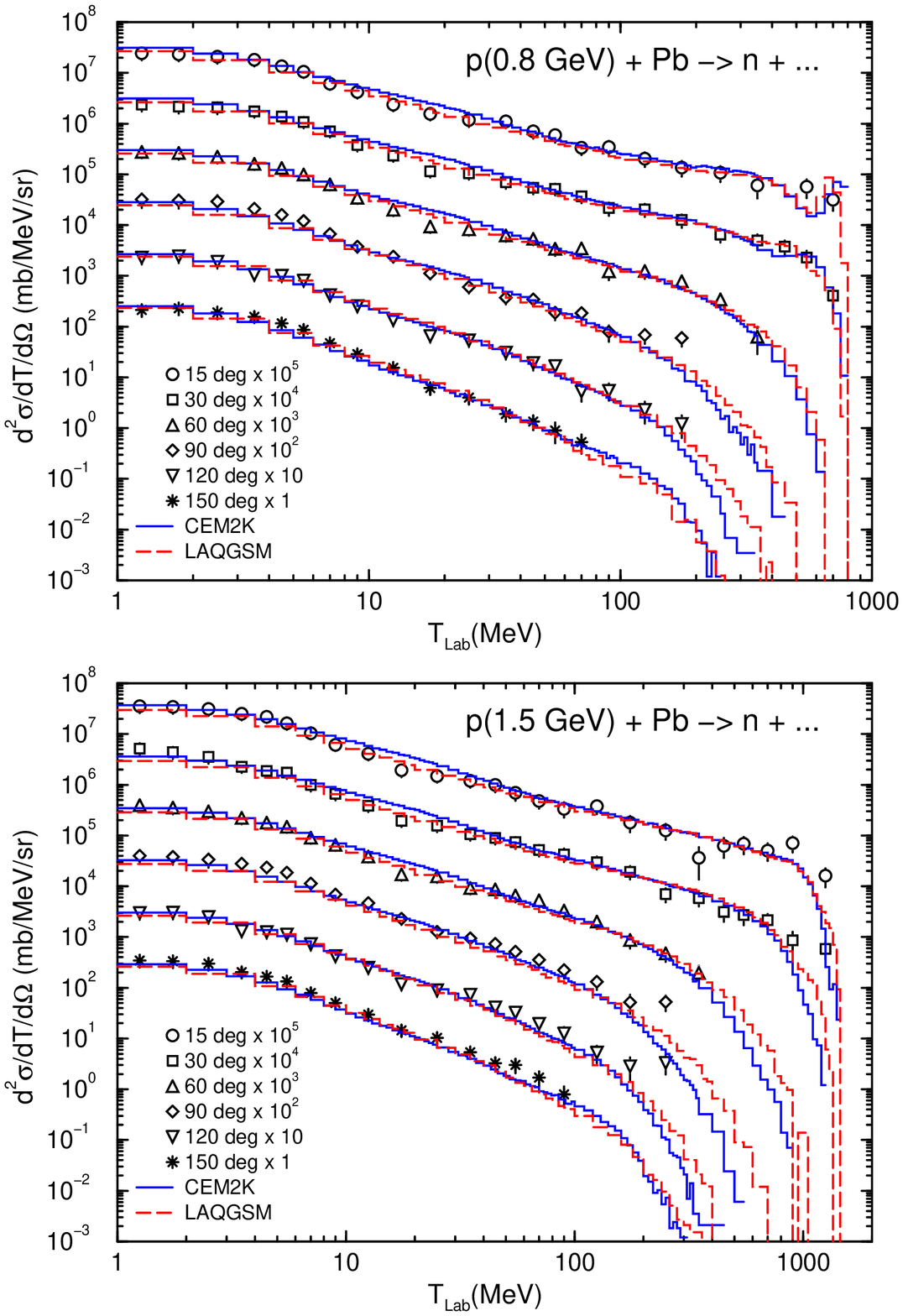}}
%\centerline{\hspace{-2.5cm} \epsfxsize 21cm \epsffile{ppbnspectrabw.eps}}
\vspace{-3.2cm}
{\bf Figure 1.}
Comparison of measured \cite{Ishibashi97} double differential cross
sections of neutrons from 0.8 and 1.5 GeV protons on Pb with CEM2k and
LAQGSM calculations.
\end{figure}

%\newpage
\begin{figure}
\vspace{-6.cm}
%\centerline{\hspace{-2cm} \epsfxsize 21cm \epsffile{ar560bw.eps}} 
\centerline{\hspace{-2cm} \epsfxsize 21cm \epsffile{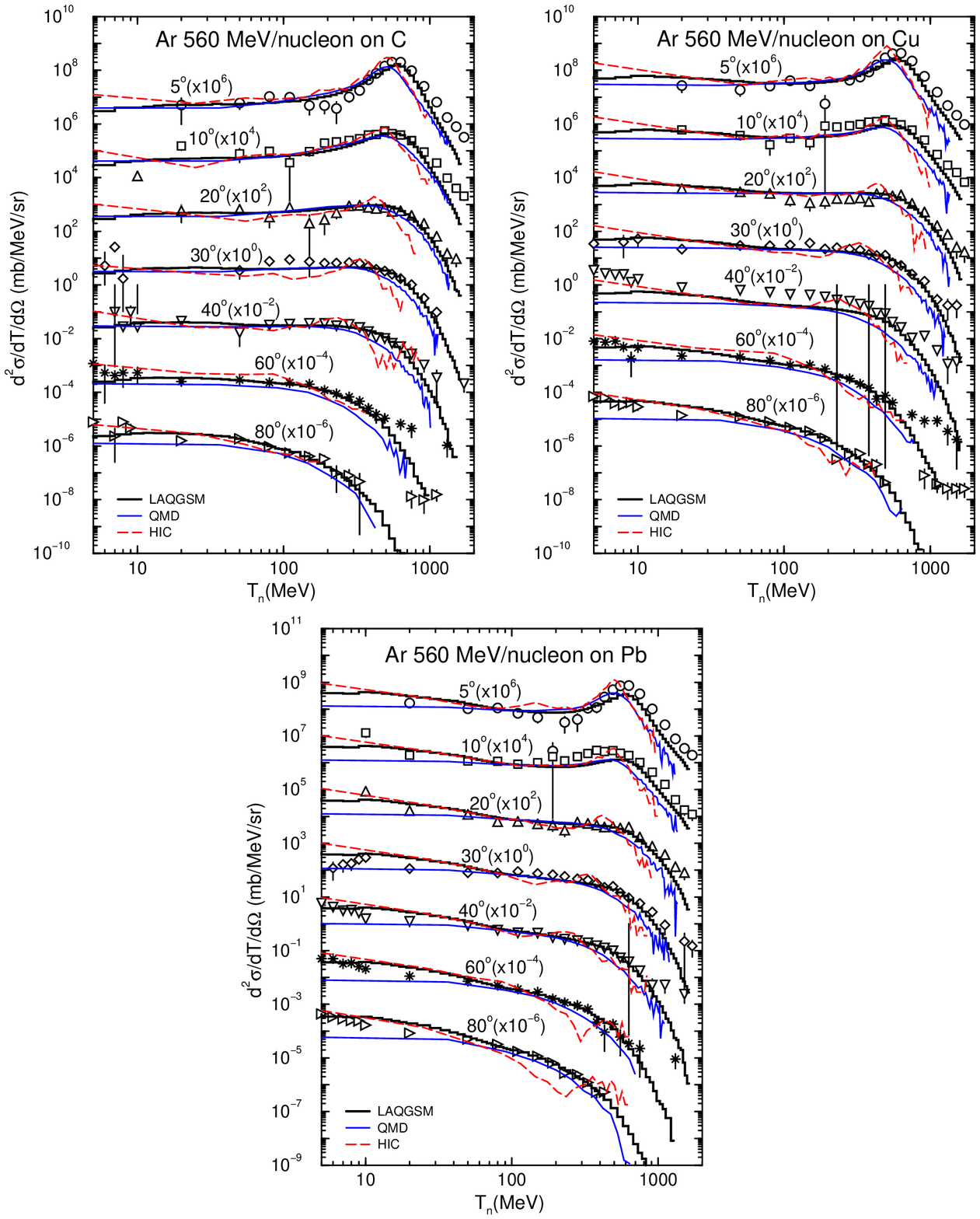}} 
\vspace{-2.5cm}
{\bf Figure 2.}
Comparison of measured \cite{Iwata01} double differential cross
sections of neutrons from 560 MeV/nucleon Ar beams on C, Cu and Pb 
with our LAQGSM results and calculations by QMD \cite{Aichelin91} 
and HIC \cite{Bertini74} from Iwata {\it et al.}
\cite{Iwata01}.
\end{figure}

\newpage

The new ingredient in GEM in comparison with LAHET which
considers evaporation of only 6 particles (n, p, d, t, $^3$He, and $^4$He)
is that Furihata included the possibility of evaporation of up to 66 types 
of particles and fragments (both in the ground and excited states)
and incorporated into GEM several sets of parameters
used to calculate inverse cross sections and Coulomb barriers
for each ejectile (we use here only default parameters of GEM2).
The 66 ejectiles
considered by GEM are:
n, p, d, t, $^{3,4,6,8}$He, $^{6-9}$Li, $^{7,9-12}$Be, $^{8,10-13}$B,
$^{10-16}$C, $^{12-17}$N, $^{14-20}$O, $^{17-21}$F, $^{18-24}$Ne,
$^{21-25}$Na, and $^{22-28}$Mg.

The fission model used in GEM is based on Atchison's model \cite{RAL,RAL1}
as implemented in LAHET \cite{LAHET},
often referred in the literature as the Rutherford Appleton Laboratory (RAL)
model, which is where Atchison developed it. There are two choices of 
parameters for the fission model: one of them is the original parameter 
set by Atchison \cite{RAL,RAL1} as implemented in LAHET \cite{LAHET}, 
and the other is a parameter set evaluated by Furihata \cite{GEM2,GEM2a},
used here as a default of GEM2.

The Atchison fission model is designed to only describe fission of
nuclei with $Z \geq 70$. It assumes that fission competes only with
neutron emission, {\it i.e.}, from the widths $\Gamma_j$ of n, p, d, 
t, $^3$He, and $^4$He,
the RAL code calculates the probability of evaporation of any 
particle. When a charged particle is selected to be evaporated, 
no fission competition is taken into account. When a neutron is
selected to be evaporated, the code does not actually simulate its 
evaporation,
instead it considers that fission may compete,
and chooses either fission or evaporation of a neutron according to
the fission probability $P_f$. This quantity is treated by the RAL code 
differently
for the elements above and below $Z=89$. The reasons Atchison split the 
calculation of the fission probability $P_f$ are: (1) there is
very little experimental information on fission in the region $Z=85$ to 88,
(2) the marked rise in the fission barrier for nuclei with $Z^2/A$ below
about 34 (see Fig.\ 2 in \cite{RAL1}) together with the disappearance of 
asymmetric 
mass splitting, indicates that a change in the character of the fission
process occurs. If experimental information were available, a split between
regions about $Z^2/A \approx 34$ would more sensible \cite{RAL1}.

1) $70 \leq Z_j \leq 88$. For fissioning nuclei with $70 \leq Z_j \leq 88$,
GEM uses the original Atchison calculation of the neutron emission
width $\Gamma_n$ and fission width $\Gamma_f$ to estimate the fission
probability as
\beq
P_f = {\Gamma_f \over {\Gamma_f+\Gamma_n}}={1 \over {1+\Gamma_n/\Gamma_f}}.
\eeq
Atchison uses \cite{RAL,RAL1} the Weisskopf and Ewing statistical model
\cite{ Weisskopf}
with an energy-independent pre-exponential factor for the level density 
(see Eq.\ (6) in \cite{SantaFe02}) and Dostrovsky's \cite{Dostrovsky}
inverse cross section for neutrons
and estimates the neutron width $\Gamma_n$ as
\bea
\Gamma_n = 0.352 \Bigl(1.68 J_0 + 1.93A_i^{1/3}J_1 
+ A_i^{2/3}(0.76J_1 - 0.05 J_0)\Bigr),
\eea
where $J_0$ and $J_1$ are functions of the level density parameter $a_n$
and $s_n \Bigl(=2\sqrt{a_n(E-Q_n-\delta)}\Bigr)$ as
$$J_0 = { {(s_n-1) e^{s_n} + 1} \over {2 a_n} } \mbox{ and } 
J_1 = { {(2s_n^2 - 6s_n + 6) e^{s_n} + s_n^2 -6} \over {8a_n^2} }.$$
Note that the RAL model uses
a fixed value for the level density parameter $a_n$, namely
\beq
a_n = (A_i - 1) /8,
\eeq
and this approximation is kept in GEM when
calculating the fission probability according to Eq.\ (1), though it 
differs from the Gilbert-Cameron-Cook-Ignatyuk (GCCI)
parameterization (see Eq. (7) in \cite{SantaFe02}) 
used in GEM to calculate particle evaporation widths.
The fission width for nuclei with $70 \leq Z_j \leq 88$ is calculated 
in the RAL model and in GEM as
\beq
\Gamma_f = { {(s_f - 1)e^{s_f} + 1} \over a_f },
\eeq
where $s_f = 2 \sqrt{a_f (E-B_f - \delta)}$ and the level density parameter
in the fission mode $a_f$ is fitted by Atchison \cite{RAL1}
to describe the measured
$\Gamma_f / \Gamma_n$ as:
\beq
a_f = a_n \Bigl(1.08926 + 0.01098 ( \chi - 31.08551)^2\Bigr),
\eeq
and $\chi = Z^2/A$.
The fission barriers $B_f$ [MeV] are estimated as
\beq
B_f = Q_n + 321.2 - 16.7 { {Z^2_i} \over A} + 0.218 
\Biggl({ {Z^2_i} \over {A_i} }\Biggr)^2 .
\eeq
Note that neither the angular momentum nor the excitation energy of the 
nucleus are taken into account in the estimate of the fission barriers.

2) $Z_j \geq 89$. For heavy fissioning nuclei with $Z_j \geq 89$
GEM follows the RAL model \cite{RAL,RAL1} and does not calculate at all
the fission width $\Gamma_f$ and does not use Eq.\ (1) to estimate
the fission probability $P_f$. Instead, the following semi-empirical
expression obtained by Atchison \cite{RAL,RAL1} 
by approximating the experimental values of
$\Gamma_n / \Gamma_f$ published by
Vandenbosch and Huizenga \cite{Vandenbosch} is used to calculate
the fission probability:
\beq
\log (\Gamma_n / \Gamma_f ) = C(Z_i) ( A_i - A_0(Z_i)),
\eeq
where $C(Z)$ and $A_0(Z)$ are constants dependent on the nuclear charge $Z$
only. The values of these constants are those
used in the current version of LAHET \cite{LAHET} 
and are tabulated in Table 5 of Ref. \cite{SantaFe02}
(note that some adjustments of these values have been done since
Atchison's papers \cite{RAL,RAL1} were published).  

The selection of the mass of the fission fragments depends on whether the
fission is symmetric or asymmetric. For a pre-fission nucleus with
$Z^2_i/A_i \leq 35$, only symmetric fission is allowed. For $Z^2_i/A_i > 35$,
both symmetric and asymmetric fission are allowed, depending on the
excitation energy $E$ of the fissioning nucleus. No new parameters were 
determined for asymmetric fission in GEM.

For nuclei with $Z^2_i/A_i > 35$, whether the fission is symmetric or not is 
determined by the asymmetric fission probability $P_{asy}$
\beq
P_{asy} = { {4870e^{-0.36E}} \over {1+4870e^{-0.36E} } } .
\eeq
For asymmetric fission, the mass of one of the 
post-fission fragments $A_1$ is selected from a Gaussian distribution of mean 
$A_f = 140$ and width $\sigma_M = 6.5$. The mass of the second fragment is
$A_2 = A_i -A_1$.

For symmetric fission, $A_1$ is selected from
the Gaussian distribution of mean $A_f = A_i/2$ and two options for the 
width $\sigma_M$ as described in \cite{GEM2,GEM2a,SantaFe02}.

The charge distribution of fission fragments is assumed to be a Gaussian 
distribution of mean $Z_f$ and width $\sigma_Z$. $Z_f$ is expressed as
$$
Z_f = { {Z_i+Z'_1 -Z'_2} \over 2} \mbox{ , where }
Z_l' = {{65.5A_l} \over {131+A_l^{2/3}}} \mbox{ , and $l=1$ or 2}.
$$
The original Atchison model uses $\sigma_Z = 2.0$. An investigation
by Furihata \cite{GEM2a} suggests that $\sigma_Z = 0.75$ provides a better 
agreement with data; therefore $\sigma_Z = 0.75$ is used in GEM2
and in all our calculations.

The kinetic energy of fission fragments [MeV] is determined by a
Gaussian distribution with mean $\eps_f$ and width $\sigma_{\eps_f}$.
The original parameters in the Atchison model are:
$$ \eps_f = 0.133Z_i^2/A_i^{1/3} - 11.4 \mbox{, and }
\sigma_{\eps_f} = 0.084 \eps_f .$$
Furihata's parameters in GEM2, which we also use, are:
$$
\eps_f =
\cases{0.131 Z^2_i/A_i^{1/3} \mbox{, for }Z^2_i/A^{1/3}_i \leq 900 , & \cr 
0.104Z^2_i/A_i^{1/3} + 24.3 \mbox{, for } 900 < Z^2_i/A_i^{1/3}\leq 1800 ,&\cr}
$$
and
$$
\sigma_{\eps_f} =
\cases{C_1 (Z^2_i/A^{1/3}_i-1000)+C_2\mbox{, for }Z^2_i/A_i^{1/3} > 1000 ,&\cr 
C_2 \mbox{, for }Z^2_i/A_i^{1/3} \leq 1000 , & \cr}
$$
where $C_1 = 5.70 \times 10^{-4}$ and $C_2 = 86.5$.
More details may be found in \cite{GEM2a}.

We note that Atchison also has modified his original version using 
recent data and published \cite{RAL98} an improved (and more complicated)
parameterization for many quantities and distributions in his model,
but these modifications \cite{RAL98} are not yet included either in LAHET or 
in GEM2.

We have merged the GEM2 code with our CEM2k and LAQGSM, initially keeping all
the default options in GEM2. We began by concentrating on an analysis of the 
recent GSI measurements in inverse kinematics as the richest and
best data set for testing this kind of model. 
As mentioned above, to understand the role of preequilibrium
particle emission, we performed calculations of all the reactions we
tested both taking into account preequilibrium particle emission and
ignoring it, {\it i.e.}, going directly to GEM2 after the intranuclear 
cascade stage of a reaction described by CEM2k or LAQGSM. 

If we merge GEM2 with CEM2k without any modifications, the new code
does not describe correctly the fission cross section (and the yields of
fission fragments) whether we take into account preequilibrium emission
(see the short-dashed 
%red 
line on Fig. 3) or not (see the long-dashed 
%blue
line on Fig. 3). Such results were anticipated, as Atchison
fitted the parameters of his RAL fission model when it was coupled 
with the Bertini INC \cite{BertiniINC} which differs
from our INC. In addition, he did not model 
preequilibrium emission. Therefore, the distributions of fissioning
nuclei in $A$, $Z$, and excitation energy $E^*$ simulated by Atchison
differ significantly from the distributions we get; 
as a consequence, all the fission characteristics are also different.

%\newpage
\begin{figure}
\vspace{-5.cm}
\centerline{\hspace{-2cm} \epsfxsize 21cm \epsffile{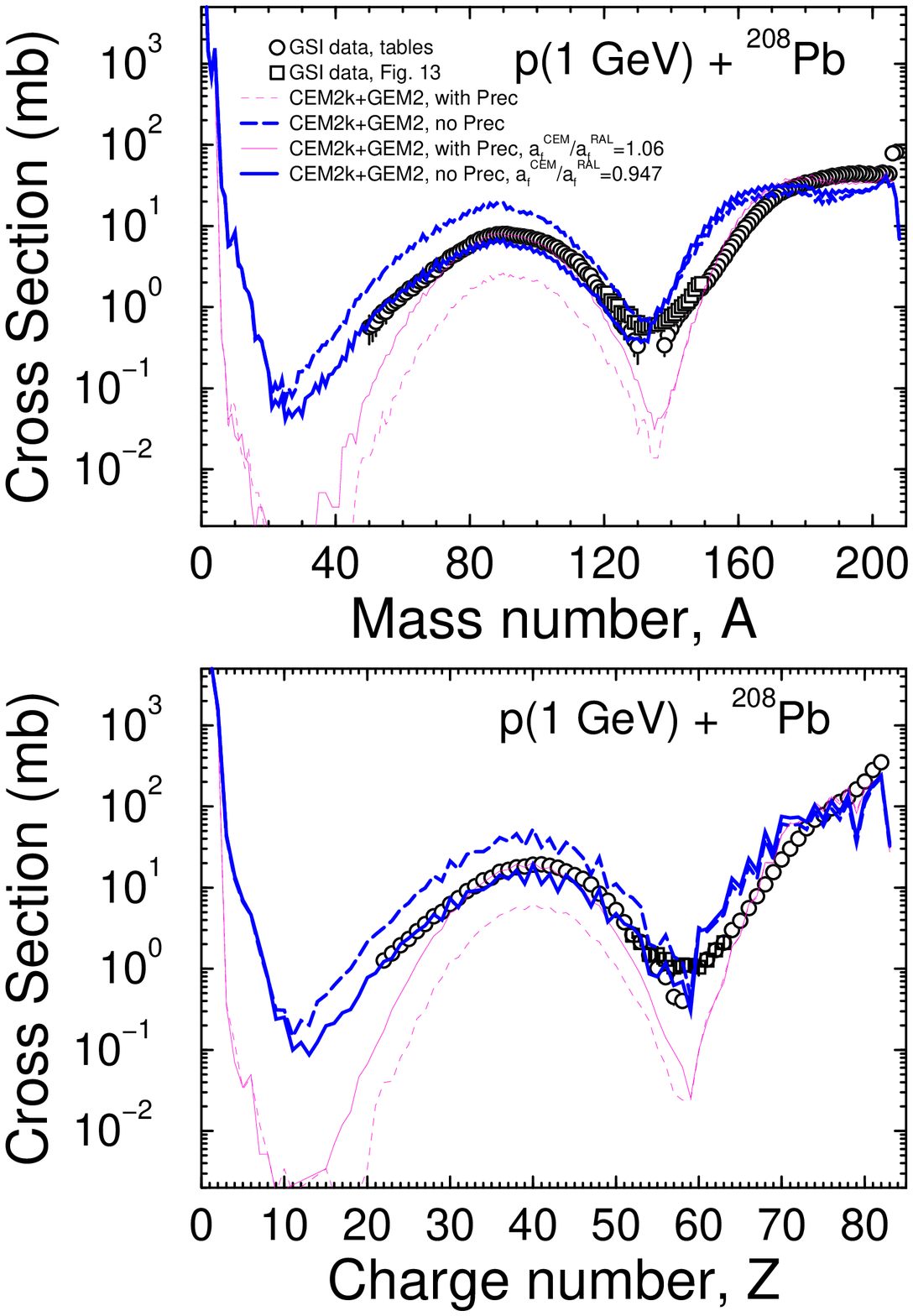}} 
%\centerline{\hspace{-2cm} \epsfxsize 21cm \epsffile{pbagem2bw.eps}} 
\vspace{-5.cm}
{\bf Figure 3.}
Comparison of the experimental
mass and charge distributions of the
nuclides produced in the reaction p(1 GeV) + Pb (circles
show data from Tabs. 3 and 4 of Ref. \cite{Enqvist01} and 
squares - from
Fig. 13 of 
Ref. \cite{Enqvist02}) 
with different calculations. 
The dashed lines show results found by merging CEM2k with GEM2 without any
modifications when preequilibrium emission is (thin lines) or is not 
(thick lines) included. Solid lines show results
from CEM2k+GEM2 with a modified $a_f$: thin lines are
for the case with preequilibrium emission ($a_f^{CEM} / a_f^{RAL} = 1.06)$
and thick lines show the results without preequilibrium emission
($a_f^{CEM} / a_f^{RAL} = 0.947)$.
\end{figure}

Furihata used GEM2 coupled either with the Bertini INC 
\cite{BertiniINC} or with
the ISABEL \cite{ISABEL} INC code, which also differs from our INC, and did 
not include preequilibrium particle emission. Therefore the
real fissioning nuclei simulated by Furihata differ from the ones in
our simulations, and the parameters adjusted by Furihata to work the best 
with her INC should not be the best for us. To get a good description 
of the fission cross section (and fission-fragment yields)
we need to modify at least one parameter in GEM2, namely to adjust
the level density parameter $a_f$ to get the correct fission cross 
section (see Eq.\ (5)), 
in the case of fissioning nuclei with $Z \leq 88$ (pre-actinides), and 
the parameter $C(Z)$ (see Eq.\ (7)) for fissioning nuclei
with $Z > 88$ (actinides). From the dashed lines on Fig.\ 5 we see that we 
need to enlarge $a_f$ in our code to get a proper fission cross section
when we include preequilibrium emission
(the excitation energy of our fissioning nuclei and their $A$ and $Z$ are
smaller than provided by the Bertini or ISABEL INC without preequilibrium),
and we need to decrease $a_f$ in the case without preequilibrium.
By increasing $a_f$ by 1.06 compared
with the original RAL and GEM2 value ($a_f^{CEM} /a_f^{RAL} = 1.06$), 
we are able to reproduce correctly with CEM2k+GEM2
the fission cross section for this reaction when we 
take into account preequilibrium emission (below, we label such 
results as ``with Prec'').
In the case with no preequilibrium emission, a proper
fission cross section is obtained for $a_f^{CEM} /a_f^{RAL} = 0.947$
(we label such results as ``no Prec''). We choose
these values for $a_f$ for all our further CEM2k+GEM2
calculations of this reaction
and do not change any other parameters. 

The solid lines in Fig.\ 3 show results with
these values of $a_f$. One can see that the ``no Prec''
version provide a good description of both the mass and charge
distributions and agrees better with the data for these characteristics
than the ``with Prec'' version
(that is not true for isotopic distributions of individual elements,
as we show below).
The ``with Prec'' version reproduces correctly the 
position of the maximum in both $A$ and $Z$ distributions and the yields
of fission fragments not too far from these maximums, but the calculated
distributions are narrower than the experimental ones.
This is again because both Atchison and Furihata fitted
their $A$ and $Z$ distributions using models without preequilibrium
emission, which provide higher values for the excitation energy,
$A$, and $Z$ of fissioning nuclei. This means that to get a good
description of  $A$ and $Z$ distributions for fission fragments using
GEM2 in CEM2k ``with Prec'', we would need to modify the 
$A$ and $Z$ distributions of fission fragments in GEM2, 
making them wider. This would take us beyond the
scope of the present work and here we do not vary any more parameters 
than we have already discussed.

Fig. 4 shows the GSI measurements \cite{Enqvist01} of the $A$ and
$Z$ distributions of the kinetic energy of products from the same reaction
compared with our CEM2k+GEM2 calculations both with and without
preequilibrium emission. Both versions of our calculations are in
reasonable agreement with the data.

Mass and charge distributions of the yields or kinetic energies of the
nuclides produced show only general trends and are not sensitive enough
to the details of a reaction. It is much more informative to study the
characteristics of individual nuclides and particles produced in a reaction.
Fig. 5 shows a comparison of the experimental data on production yields 
of thirteen separate isotopes with $Z$ lying from 22 to 82 from the same 
reaction measured at GSI \cite{Enqvist01} with our calculations using both the
``with Prec'' (upper plot) and ``no Prec'' (middle plot) versions.

%\newpage
\begin{figure}
\vspace{-5.cm}
\centerline{\hspace{-2cm} \epsfxsize 21cm \epsffile{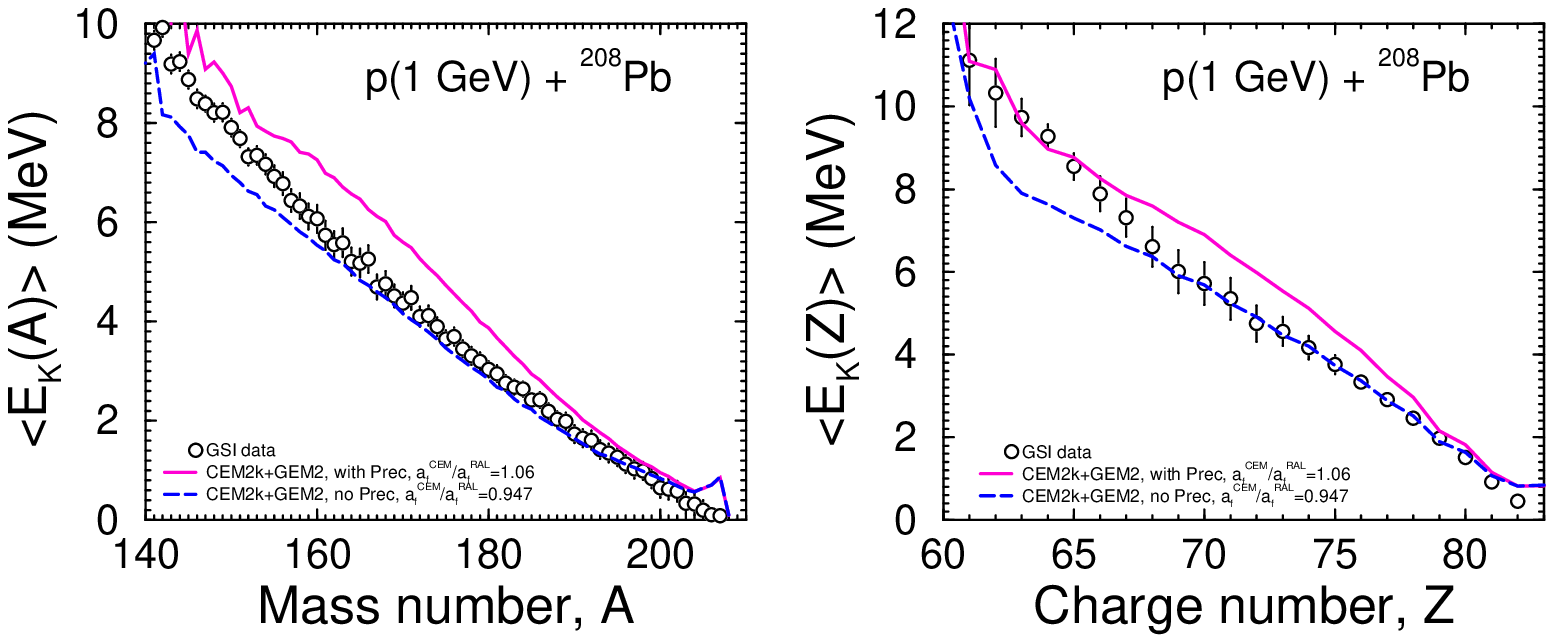}} 
%\centerline{\hspace{-2cm} \epsfxsize 21cm \epsffile{ppbtkinbw.eps}} 
\vspace{-19.cm}
{\bf Figure 4.}
Comparison of the experimental \cite{Enqvist01}
mass and charge distributions of the
spallation-residue
kinetic energies of the 
nuclides produced in the reaction p(1 GeV) + Pb (circles) with
our CEM2k+GEM2 calculations: ``with Prec'' results are shown
by solid lines, ``no Prec'' results are shown by dashed lines.
\end{figure}

The agreement (or disagreement) of our calculations with these data
is different from what we have for the integral
$A$ or $Z$ distributions in Figs. 3
and 4: We see that for the isotopes produced in the spallation
region (not too far from the target) and for fission fragments in
the region with the maximum yield, the version ``with Prec'' agree
much better with the data than the version ``no Prec''.
Only for production of isotopes at the border between spallation
and fission and between fission and fragmentation does
the version ``with Prec'' underestimates the data, due to too narrow
$A$ and $Z$ distributions in the simulation of fission fragments, as
we discussed previously. The ``no Prec'' version agrees better with
the data in these transition regions but are in worse
agreement for isotopes both in the spallation region and in the
middle of the fission region. We conclude that if a model agrees well
with some $A$ or $Z$ distributions it does not necessarily mean that 
it also describes well production of separate isotopes.
In other words, integral $A$ and $Z$ distributions
are not sensitive enough to develop and test such models, 
a practice which is often used in the literature.

The lower plot in Fig. 5 shows results of calculations with a 
version of CEM2k+GEM2 with reduced preequilibrium emission. We
prefer to discus this version together with results by LAQGSM+GEM2,
and will return to this plot later.

It is more difficult for any model to describe correctly the
energy dependence for the production cross sections
of different isotopes, {\it i.e.,} excitation functions.
We calculated using both the ``with Prec'' and ``no Prec'' versions
of CEM2k+GEM2 all the excitation functions for the same reaction, p + Pb,
for proton energies from 10 MeV to 3 GeV and compared our results with
all available data from our compilation referred to here as T-16 Library
(``T16 Lib'') \cite{T16Lib}. Only several typical examples from our comparison
are shown below.

Fig. 6 shows two examples of excitation functions for the production
of several isotopes in the spallation (first two columns of plots)
and fission (the last two columns of

%\newpage
\begin{figure}
\vspace{-5.cm}
\centerline{\hspace{-0.1cm} \epsfxsize 20cm \epsffile{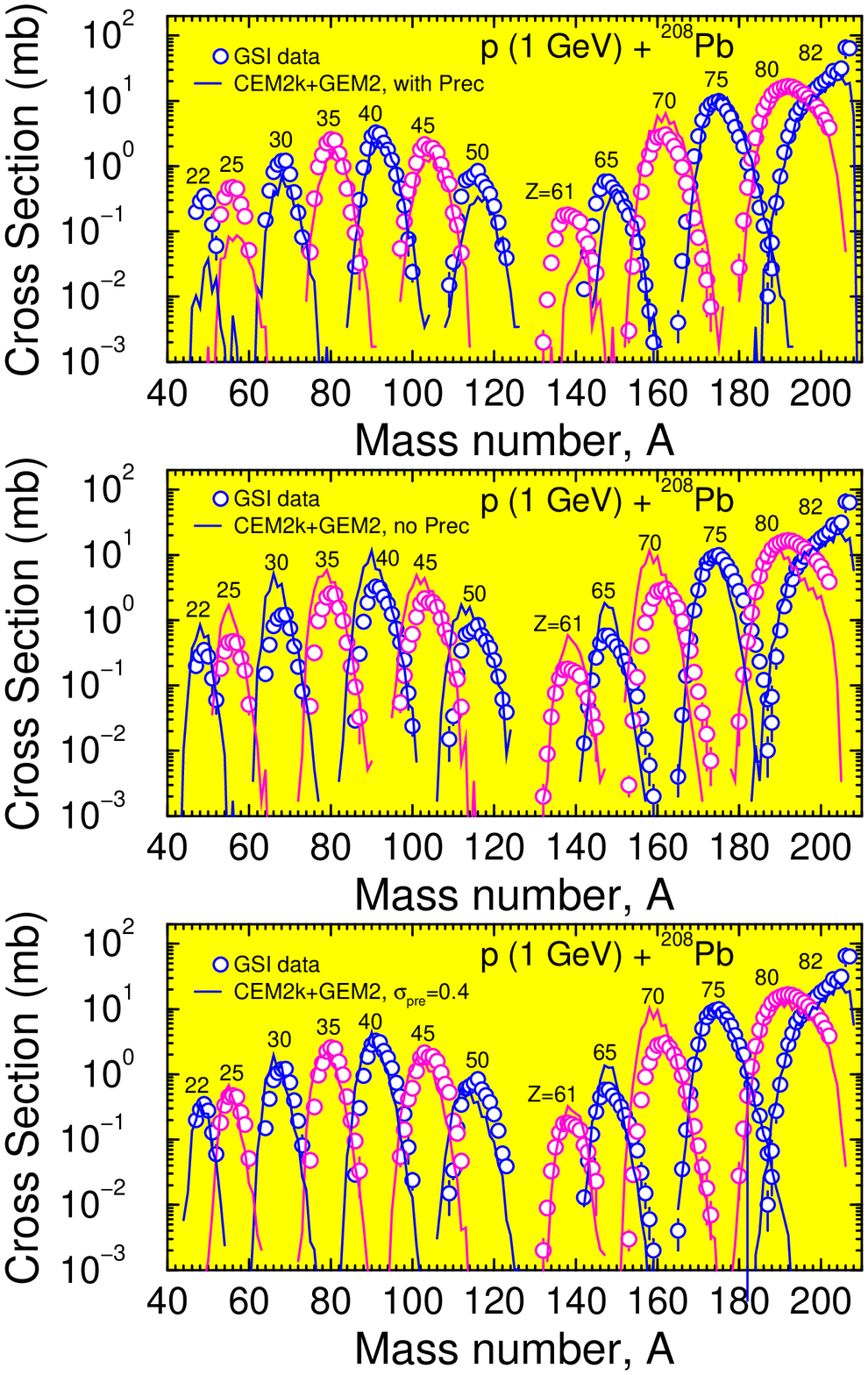}} 
%\centerline{\hspace{-0.1cm} \epsfxsize 20cm \epsffile{2282satifbw.eps}} 
\vspace{-2.5cm}
{\bf Figure 5.}
Experimental \cite{Enqvist01} mass distributions of the cross sections 
of thirteen isotopes with the charge $Z$ from 22 to 82
compared with our CEM2k+GEM2 calculations. ``With Prec'' 
results are shown on the upper plot, ``no Prec'' results 
are shown in the middle, while results with reduced preequilibrium
emission according to Eq. (9), in the lower one.
\end{figure}

%\newpage
\begin{figure}
\vspace{-5.cm}
\centerline{\hspace{-0.7cm} \epsfxsize 19.7cm \epsffile{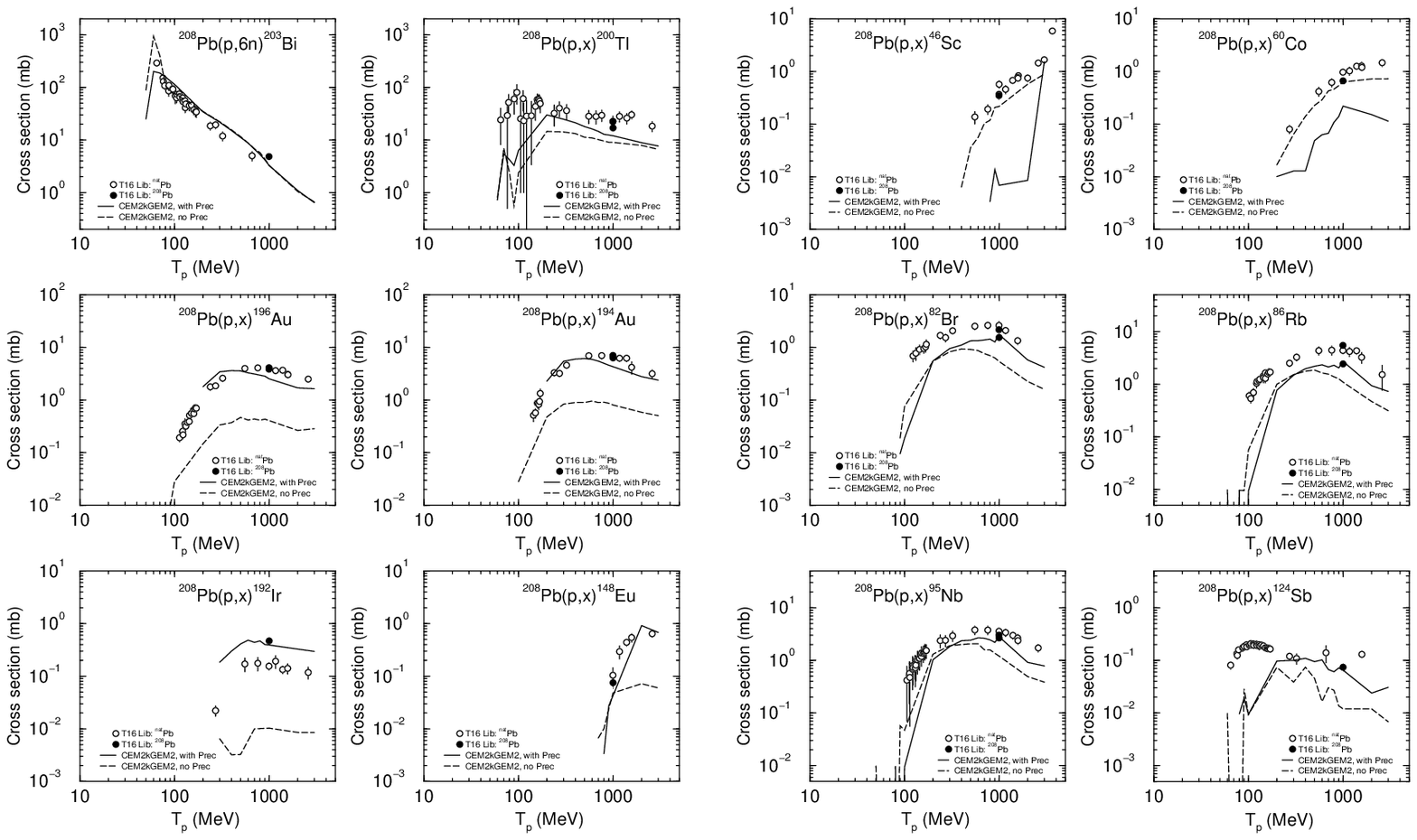}} 
\vspace{-14.5cm}
{\bf Figure 6.}
Excitation functions for the production of $^{203}$Bi, $^{200}$Tl,
$^{196}$Au, $^{194}$Au, $^{192}$Ir, $^{148}$Eu, 
$^{124}$Sb, $^{95}$Nb, $^{86}$Rb, $^{82}$Br, 
$^{60}$Co, and $^{46}$Sc
from p+$^{208}$Pb. Results by CEM2k+GEM2
``with Prec'' are shown by solid lines and ``no Prec'' by dashed lines.
Experimental data (filled circles for $^{208}$Pb targets and
opaque circles for $^{nat}$Pb) are from our LANL compilation 
(``T16 Lib") \cite{T16Lib} and are available from the authors upon request.
\end{figure}

{\noindent
plots regions).}
One can see a not too good 
but still reasonable agreement of both calculations with many data
(note that most of the data were measured for $^{nat}$Pb targets, 
while our calculations were done for $^{208}$Pb). 
We see that merging CEM2k with GEM2 allows us to
reasonably describe yields of fission fragments, while in the old standard
CEM2k we do not have any fission fragments and are unable to describe
such reactions at all.
We see that as shown in Fig.\ 5  for
a single proton energy of 1 GeV, the ``with Prec'' version
agrees better with the data in the whole energy region 
both for spallation products and 
for the production of most of the fission fragments.
Only on the border between fission and fragmentation regions ($^{46}$Sc
and $^{60}$Co) does the ``no Prec'' version agree much better with
the data than the ``with Prec'' version; the reason for this we have
already discussed.
Similar results 
were obtained for excitation functions of many other isotopes in the 
spallation and fission regions.
On the whole, the version ``with Prec'' reproduces most of the experimental 
excitation functions better that the version ``no Prec''. 

In Fig. 7 we show examples of excitation functions for the production
of light fragments (the first two columns of plots), 
in the fragmentation region, that are produced in 
CEM2k+GEM2 only via evaporation (the contribution to the yield of these
isotopes from fission or deep spallation is negligible), and of nucleons
and complex particles up to $\alpha$ (the last two columns of plots).
We see that with the ``no Prec'' version, GEM2 reproduces correctly the
yields of light fragments $^{6}$He, $^{9}$Li,  $^{7}$Be, $^{13}$N,
and $^{18}$F, and not so well the
excitation functions for heavier fragments like $^{22}$Na.
With increasing mass of the fragment, the calculations progressively
underestimate their yields. Note, that in  \cite{SantaFe02},
we got very 
similar results for excitation functions for the
p+Au reaction. 
The version 

\newpage
\begin{figure}[!ht]
\vspace{-5.cm}
\centerline{\hspace{-0.7cm} \epsfxsize 19.7cm \epsffile{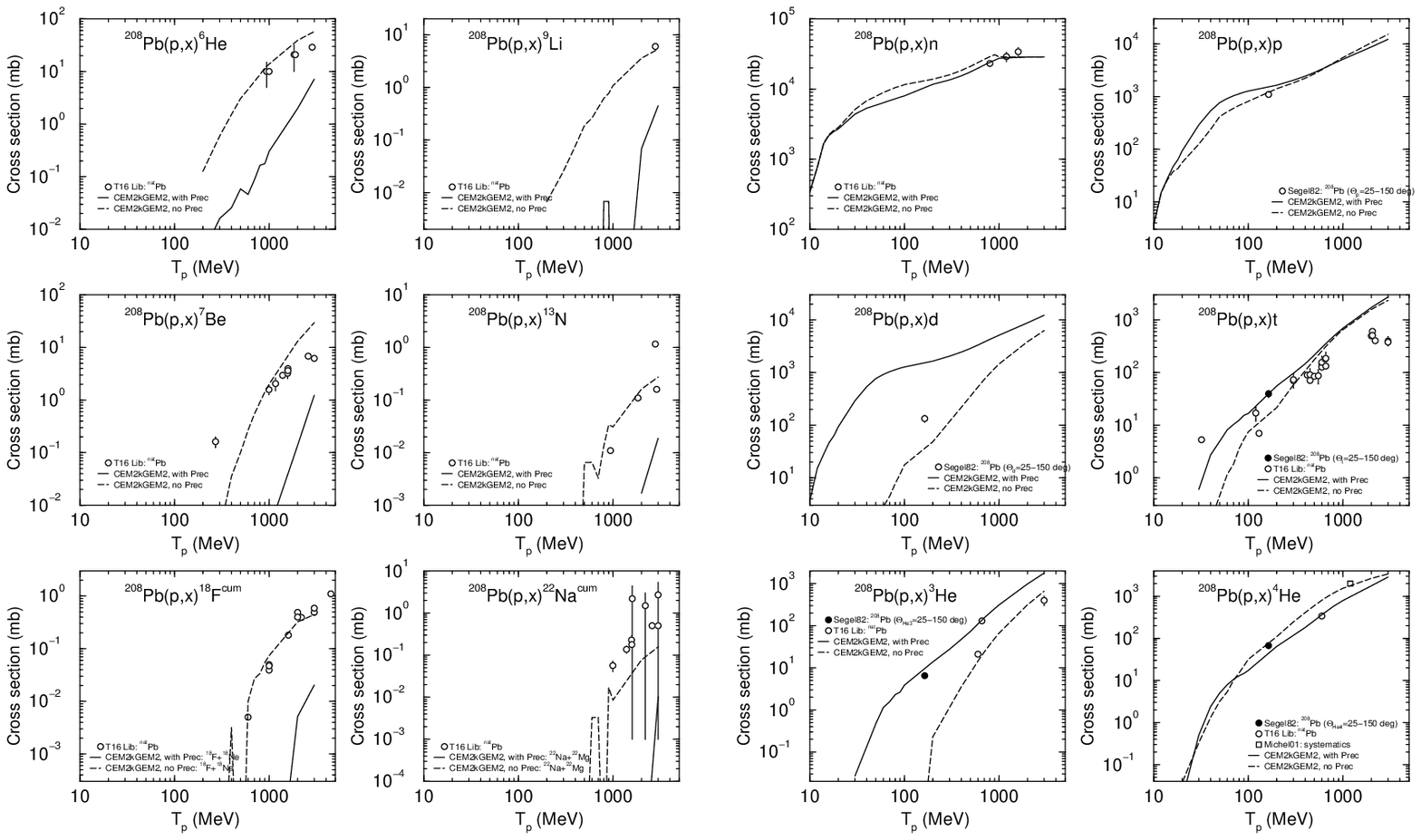}} 
\vspace{-14.5cm}
{\bf Figure 7.}
The same as Fig.\ 6 but for the production of
$^{22}$Na, $^{18}$F,
$^{13}$N, $^{7}$Be, $^{9}$Li, $^{6}$He, 
$^{4}$He, $^{3}$He, t, d, 
p, and n.
The complex particles up to $^4$He
are produced in CEM2k+GEM2 via evaporation and preequilibrium emission;
n and p are also produced during the cascade stage; fragments
heavier than $^4$He are produced only via evaporation. Data labeled
as ``Segel82" are from Ref. \cite{Segel82}.
\end{figure}

{\noindent
``with Prec'' strongly underestimates the yields 
of all these fragments,
and this is again not surprising, 
as Furihata developed her model
and fitted all parameters without taking into account preequilibrium
processes.} 

Undeniably, the parameters determining the yields of evaporated
fragments in GEM (inverse cross sections and Coulomb barriers)
could be adjusted to get a good agreement with the data
for the yields of light fragments with the version ``with Prec''
(see, {\it e. g.}, how Furihata and Nakamura addressed this problem
in \cite{GEM2mod} for their version of a code without preequilibrium
emission).
This is not the aim of our present work and we will not 
do this here. Even if we were to
do this, we expect in advance to get similar results to those we got for
the ``no Prec'' version: It would be possible to
describe correctly the yields and spectra of light fragments but not of heavy
fragments like $^{24}$Na and $^{28}$Mg. To describe such heavy fragments 
(not only their yields, but also their spectra)
the model would need to be improved further, by considering 
other mechanisms for heavy fragment production in addition to the
evaporation process taken into account by GEM2.

Finally, the two last columns of plots in Fig.\ 7 show 
excitation functions for emission of nucleons 
and complex particles up to $\alpha$ for this reaction. 
Note that the data for these excitation functions are not so extensive 
and precise as we have for heavier
products: many data points were obtained by integration (plus extrapolation)
of the spectra of  particles measured only at several angles and only 
for a limited range of energy.
But even from a comparison with
these sparce and imprecise data we see that the ``with Prec'' version
describes these excitation function better than the ``no Prec''
version, just as we found in \cite{SantaFe02}
for the p+Au reaction. This is an expected result as the high Coulomb
barriers for 
heavy nuclear targets oppose evaporation of low energy
charged particles and the main contribution to their yields comes
from preequilibrium emission from highly excited
pre-compound nuclei.

For completeness sake, we show here also an example of results 
from a calculation with the merged
CEM2k+GEM2 code of a reaction on an actinide, p(100 MeV) + $^{238}$U.
Generally,
to get for actinides a proper fission cross section, we need to adjust
in GEM2 the parameters $C(Z)$ (or, also $A_0(Z)$) in Eq.\ (7), as they
were fitted by Atchison to work the best with Bertini's INC and we have
in CEM2k our own INC. As mentioned above, for actinides, Eq.\ (1) is not
used in GEM2 and $a_f$ is not used in any calculations, 
therefore we do not need to adjust $a_f/a_n$,
for fissioning nuclei with $Z > 88$. 
We found that 
for this particular reaction, p(100 MeV) + $^{238}$U, 
we get
with CEM2k+GEM2 a fission cross section in agreement with the
data without any adjustments of the parameter $C(Z)$ in GEM2,
i.e., we can use just the default parameters of GEM2.
Nevertheless, 
our results for other reactions show that
for higher energies of the incident protons 
or for other target-nuclei,
the parameter $C(Z)$ 
has to be fitted to get a correct fission cross section
when GEM2 is coupled either with CEM2k or with LAQGSM.
In addition, we should mention that for reactions on actinides at 
intermediate or high energies, the parameter
$a_f^{CEM} / a_f^{RAL}$ should also be fitted along with
$C(Z)$. In some simulated events several protons
can be emitted at the cascade and preequilibrium stages of the reaction,
as well as at the evaporation stage, before the compound
nucleus actually fissions
(also complex particles can be emitted before fission), 
and the charge of the fissioning nucleus can 
have $Z \le 88$, even when the initial
charge of the target has $Z > 88$. At the same time, for $Z \leq 88$,
due to charge exchange reactions, the charge of the fissioning nucleus may
exceed 88, so that we would need to fit as well $C(Z)^{CEM}/C(Z)^{RAL}$.
This is a peculiarity of treating the fission 
probability $P_f$ differently for the elements above and below $Z=89$ 
in the Atchison model.

Fig. 8 shows mass distributions of products from 
p(100 MeV) + $^{238}$U calculated with both versions of CEM2k+GEM2
compared to the available experimental data \cite{Titarenko99}
and with results by the
phenomenological code CYF of Wahl \cite{Wahl01} (short dashed lines).
We need to mention that these data are not as good for testing and 
developing models as are the GSI data measured in inverse kinematics
for the p + Pb reaction discussed above: All the data shown in Fig.\ 8
were obtained by the $\gamma$-spectrometry method. Only some of the
produced isotopes were measured, and most of the data were measured
for the cumulative yields. To get the ``experimental'' A-distribution, we
summed for each $A$ the available data taking care to not sum the individual
cross sections already included in some cumulative yields;
but the resulting A-distribution is still not complete, as many 
isotopes were not measured. This means that some theoretical values 
can be above the experimental data (where some isotopes were not 
measured) without necessarily implying disagreement between calculations 
and measurements.

One can see that both the
``with Prec'' and ``no Prec'' versions of CEM2k+GEM2
describe equally well the mass distribution of the
products from this reaction; therefore it is not possible
to choose between either of the versions from a comparison with these
not very informative data. 
We see that results by the
phenomenological code CYF of Wahl \cite{Wahl01} also agree well
with the A-distribution of measured fission products.
It is more

%\newpage

\begin{figure}[!ht]

\vspace{-4.cm}
\hspace{-19mm}
\includegraphics[width=14cm,angle=-90]{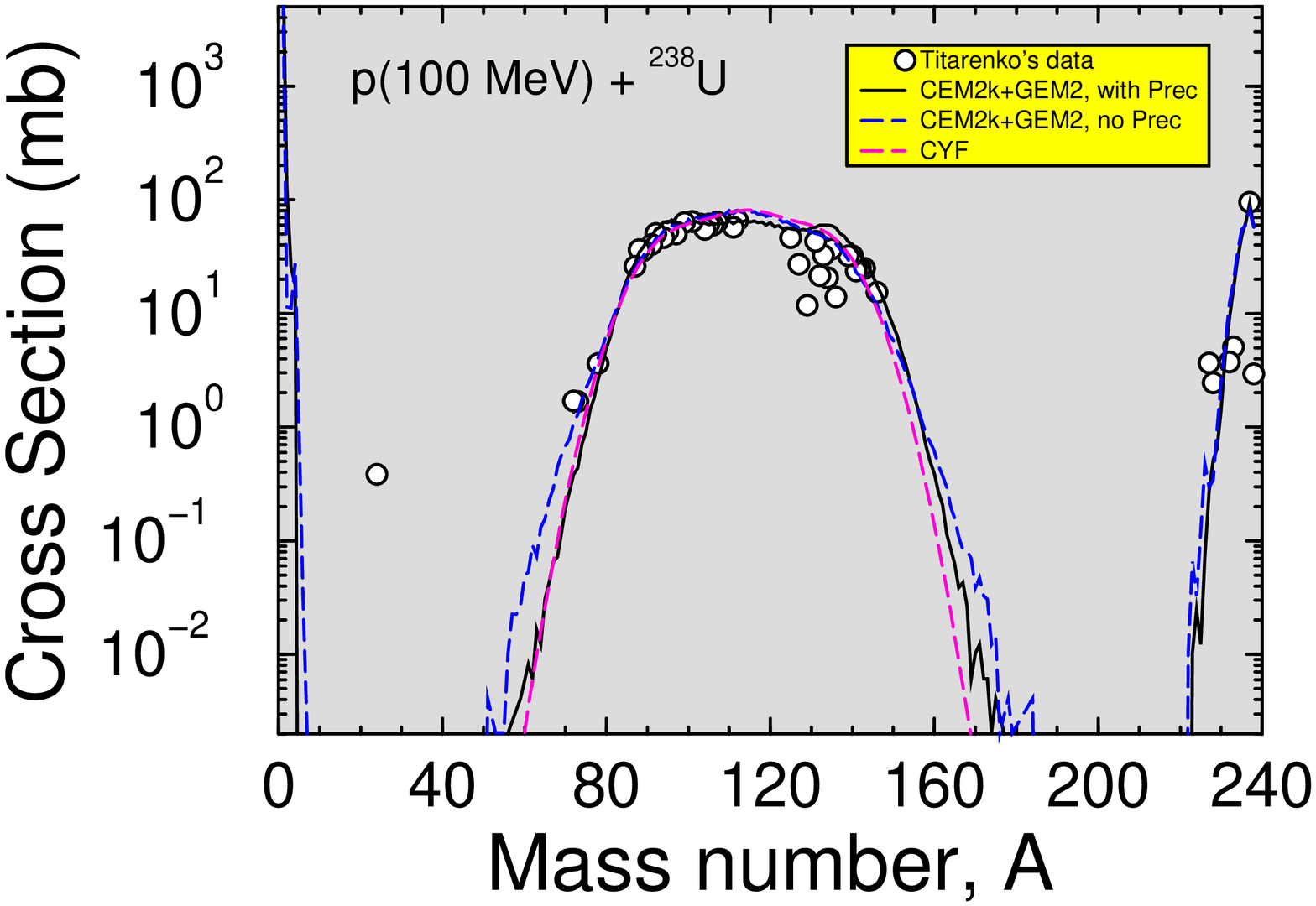}
{\bf Figure 8.}
Comparison of the experimental \cite{Titarenko99} mass
distribution of the nuclides produced in the reaction  
p(100 MeV) + $^{238}$U (circles) with calculations by the CEM2k+GEM2
model when preequilibrium  emission is
(solid lines) or is not (long dashed lines) included  and 
with results by the
phenomenological code CYF of Wahl \cite{Wahl01} (short dashed lines).
Note that  measurements were done on $^{nat}$U while calculations were
performed for $^{238}$U.
\end{figure}

{\noindent
useful to compare calculations
with the yields of individually measured isotopes.}
In Fig. 9, we compare our calculations with all cross sections measured
by Titarenko \cite{Titarenko99}
for each nuclide separately, where we can compare our
results with the data (we do not include in our comparison the nuclides
measured only either in their isomer or ground states, as our model does not
provide such information: CEM2k+GEM2 provides only yields for the sum
of isotope production cross sections both in their ground and excited states). 
We see that on the whole, the ``with Prec'' version agrees better
with most of the individually measured cross sections than the ``no
Prec'' version and for many  of the measured isotopes 
the disagreement is less than a factor of two,
with several of our calculated points coinciding with the data within
the plotting accuracy.
Nevertheless, for several isotopes
like $^{72}$Ga, $^{227}$Th$^{c}$,  $^{228}$Pa$^{c}$, and $^{238}$Np, 
we see some big disagreements.
For comparison, we also show in Fig. 9 calculations by the phenomenological
code YIELDX of Silberberg, Tsao, and Barghouty
\cite{YIELDX} 
and using its 2000 updated
version, YIELDX2k \cite{YIELDX2k},
and with the 
phenomenological code CYF by Wahl$^{65}$ often used in applications.
We see that these phenomenological systematics fail
to describe the production of many isotopes from this reaction, 
indicating that we cannot rely on phenomenological systematics and 
must develop reliable models to be used in applications.

%\newpage
\begin{figure}

\vspace{-5.cm}
\centerline{\hspace{-0.7cm} \epsfxsize 20cm \epsffile{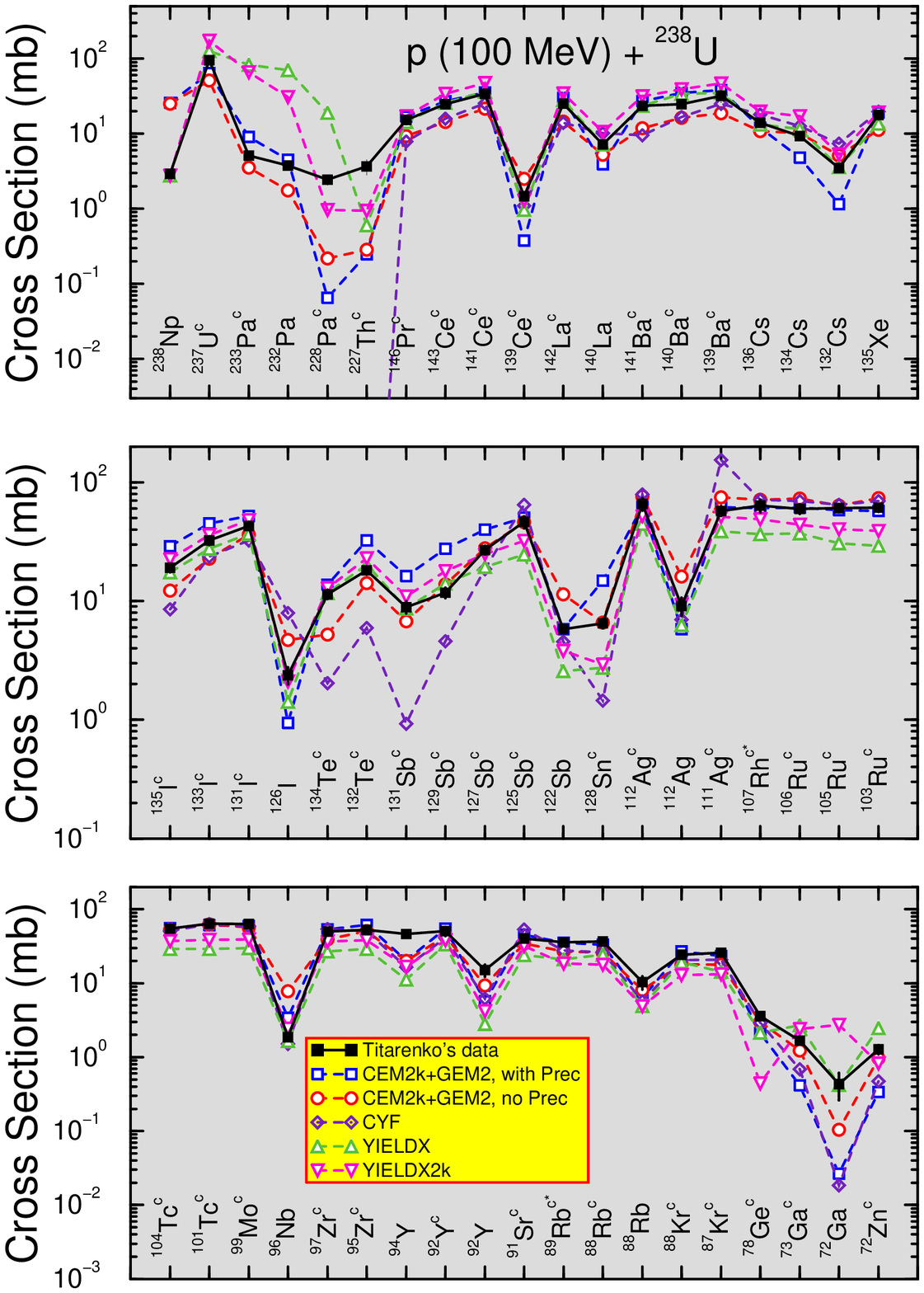}} 
%\centerline{\hspace{-0.7cm} \epsfxsize 20cm \epsffile{ubw.eps}} 
\vspace{-4.cm}
{\bf Figure 9.}
Detailed comparison between experimental \cite{Titarenko99} (filled
squares) and calculated cross sections of all individual
and cumulative (labeled with a ``c") products measured for
the reaction  
p(100 MeV) + $^{238}$U.
Our CEM2k+GEM2 ``with Prec'' results are shown by connected opaque squares
and the ``no Prec'' results are shown by connected circles.
Predictions by the phenomenological code CYF \cite{Wahl01} are shown
with connected diamonds, and results calculated by
the phenomenological systematics YIELDX \cite{YIELDX} are shown by 
connected triangles up, and using its 2000 updated
version, YIELDX2k \cite{YIELDX2k}, by connected triangles down. 
\end{figure}

Very similar results for the studied proton-induced reactions
with the ones shown above by CEM2k+GEM2
were obtained also by merging LAQGSM with GEM2. We will not discuss
such them here; instead, we will apply the merged 
LAQGSM+GEM2 code to 
analyze reactions on the same $^{208}$Pb-target (projectile,
in the case of the inverse kinematics of the GSI 
measurements \cite{Enqvist02}),
but induced by deuterons.

A- and Z-distributions of the nuclides produced in the d(1 GeV/nucleon) + Pb
reaction calculated with the LAQGSM+GEM2 code are compared in Fig. 10
with the GSI data \cite{Enqvist02}. As in case of proton-induced
reactions, we get a correct fission cross section
only by adjusting the ratio of level density parameters in the fission
and evaporation channels $a_f/a_n$ in Eq.(5), or, equivalently,
by adjusting $a_f^{CEM} / a_f^{RAL}$, if we use $a_f^{RAL}$ provided
by Eq. (5). We get a correct fission cross section for 
$a_f^{CEM} / a_f^{RAL} = 1.15$ in the ``with Prec" case (short-dashed
lines in Fig. 10) and for
$a_f^{CEM} / a_f^{RAL} = 0.97$ in the case of ``no Prec" 
(long-dashed lines in Fig. 10) when
calculating with LAQGSM+GEM2
both p- and d-induced reactions on Pb at 1 GeV/nucleon.
Note that these values differ slightly from the ones obtained above
for the CEM2k+GEM2 code (see Fig. 3). 
This can be easy understood, as the INC of
LAQGSM differs from the INC of CEM2k, therefore the nuclei simulated
by LAQGSM+GEM2 and CEM2k+GEM2 to really fission after the cascade
and preequilibrium stages of reaction and after evaporation of several 
particles from compound nuclei before fission have slightly different
average values of A, Z, and $E^*$.

The agreement (or disagreement) of the LAQGSM+GEM2 results with these
data is similar to what we get for the p+Pb reaction: the ``no Prec" version
provides a good description of both mass and charge distributions and
agrees better with the data for these characteristics than the
``with Prec" version. These results, together with the ones discussed
above for p+Pb reactions, as well as results for many other p+A
and A+A reactions measured at GSI at about 1 GeV/nucleon and analyzed
with our codes suggest us that we need to take into account preequilibrium
processes, but we need less emission of preequilibrium particles than
provided by out standard CEM2k and LAQGSM codes. We address this
problem following Veselsky \cite{sigpre}. It is assumed that the ratio
of the number of quasiparticles (excitons) $n$ at each preequilibrium
reaction stage to the number of excitons in equilibrium configuration
$n_{eq}$, corresponding to the same excitation energy, to be the
crucial parameter for determining of the probability of preequilibrium
emission $P_{pre}$. This probability for a given preequilibrium
reaction stage is evaluated using the formula
\beq
P_{pre}(n/n_{eq}) = 1 - \exp \Bigl( - { {(n/n_{eq} -1)} 
\over {2\sigma_{pre} ^2} }\Bigr)
\eeq
for $n \leq n_{eq}$ and equal to zero for $n > n_{eq}$. 
The basic assumption leading to Eq. (9) is that $P_{pre}$ depends
exclusively on the ration $n/n_{eq}$ as can be deduced from the
results of B\"{o}hning \cite{Bohning} where the density of particle-hole
states is approximately described using

%\newpage
\begin{figure}[!ht]

\vspace{-5.cm}
\centerline{\hspace{-0.4cm} \epsfxsize 22.5cm \epsffile{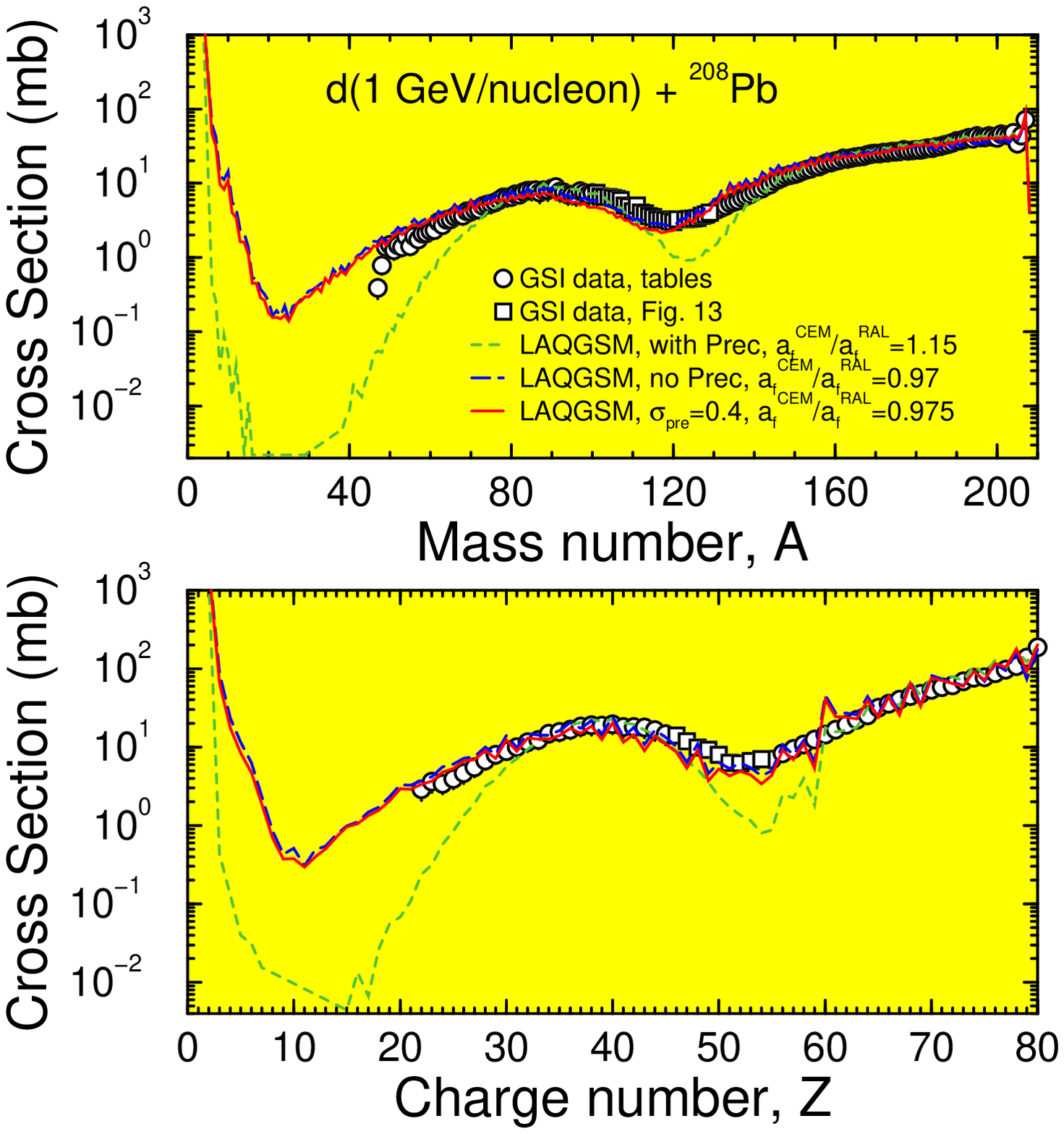}} 
%\centerline{\hspace{-0.4cm} \epsfxsize 22.5cm \epsffile{dpbbw.eps}} 
\vspace{-11.5cm}
{\bf Figure 10.}
Comparison of the experimental
mass and charge distributions of the
nuclides produced in the reaction d(1 GeV/nucleon) + Pb (circles
show data from Tabs. 2 and 3 and 
squares - from
Fig. 13 of 
Ref. \cite{Enqvist02}) 
with different calculations. 
The short-dashed lines show results found by merging LAQGSM with GEM2 
when preequilibrium emission is included
and $a_f^{CEM} / a_f^{RAL} = 1.15$,
the long-dashed lines
show the results without preequilibrium emission for
$a_f^{CEM} / a_f^{RAL} = 0.97$,
and the solid lines show the results with reduced preequilibrium emission
using $\sigma_{pre}=0.4$ and $a_f^{CEM} / a_f^{RAL} = 0.975$; 
see text for details.
\end{figure}

{\noindent
a Gaussian centered at $n_{eq}$.
The parameter $\sigma_{pre}$ is a free parameter and no dependence on
excitation energy is assumed \cite{sigpre}. Our calculations of
several reactions using different values of $\sigma_{pre}$ show that
an overall reasonable agreement with available data can be obtained
using $\sigma_{pre} = 0.4$ or 0.5 (see Fig. 11). We choose the fixed value
$\sigma_{pre} = 0.4$ both for CEM2k+GEM2 and LAQGSM+GEM2 for all
our further calculations.
}

Results with reduced preequilibrium emission according to Eq. (9) 
using $\sigma_{pre} = 0.4$ are shown in Fig. 10 with solid lines
(and in the lower plot of Fig. 5 for p+Pb reactions);
they agree better with the data than either ``with Prec" or ``no Prec"
standard results.

%\vspace{-1.5cm}
Fig. 11 shows 
a comparison of the experimental data on production yields 
of ten separate isotopes with $Z$ lying from 35 to 82 from the same 
reaction measured at GSI \cite{Enqvist02} with our LAQGSM+GEM2
calculations using both the
``with Prec'' and ``no Prec'' (upper plot) versions,
and when calculated with the version of reduced preequilibrium
emission according to Eq. (9) (lower plot). One can see that as
in the case of the p+Pb reaction (see lower plot in Fig. 5), calculations
with reduced preequilibrium emission agree much better with experimental
data than either standard ``with Prec" or ``no Prec" versions.
Similar results were obtained for all other p+A and A+A reactions
measured at GSI and calculated by our codes. Therefore we choose
the version with reduced preequilibrium emission with a fixed value
of $\sigma_{pre}$ of 0.4 as our ``standard" version both for
CEM2k+GEM2 and LAQGSM+GEM2, and all our further calculations will
be done in this approach.

Fig. 12 shows a comparison of the experimental \cite{Enqvist02}
A-distribution of the mean kinetic energies of the spallation-residue
products from the same reaction with different calculations by
LAQGSM+GEM2. One can see that for this particular reactions all
calculations agree well with the data, with a slightly better
agreement for the version with reduced preequilibrium emission.
Similar results were obtained for the Z-distribution of products
from this reaction.

Finally, for completeness sake, the last four figures (Figs. 13--16)
show a comparison of all cross sections measured at GSI for the
production of nuclides both from spallation and fission reactions
from interaction of $^{208}$Pb beams with p and d targets compared with
out ``standard" LAQGSM+GEM2 results. On can see a very good overall
agreement, taking into account that we use here our ``standard" version
of the code and no further fitting of any parameters is done.
Similar results were obtained with standard
versions of  CEM2k+GEM2 and LAQGSM+GEM2 for many other p+A and
A+A reactions measured recently at GSI. Such results will be presented
in a separate publication.
\\

{\noindent \bf Further Work} \\

Merging the  Generalized Evaporation Model code GEM2
by Furihata \cite{GEM2,GEM2a}
with our CEM2k and LAQGSM codes allows us to describe
reasonably well many fission and fragmentation reactions 
in addition to the spallation reactions already described 
well by CEM2k and LAQGSM. But to do so, ones need to
provide first a correct description of the fission cross
sections by fitting in GEM2 the ration $a_f/a_n$ 
for pre-actinides and/or the parameter C(Z) in Eq. (7) for actinides.
This is not a serious problem and we have derived already
new approximations for $a_f/a_n$ and for parameter
C(Z) in Eq. (7) that provide for CEM2k+GEM2
and LAQGSM+GEM2 correct fission cross section calculation
both for pre-actinides and actinides for a large range of
projectile energies. Such new approximations will be presented
together with examples of our results in a separate publication.

Some reactions, like production
of fission fragments at the borders between fission and fragmentation 
or between fission and emission of heavy fragments
like Na and Mg are poorly described by CEM2k+GEM2 and LAQGSM+GEM2
in their current versions. 
This disagreement does not discourage us;
the results of the present work suggest that some
of the fission and evaporation parameters of GEM2 can be adjusted
to get a much 
better description of all reactions.
This lends credibility to such an approach. 

%\newpage
\begin{figure}
\vspace{-6.cm}
\centerline{\hspace{-0.4cm} \epsfxsize 22.5cm \epsffile{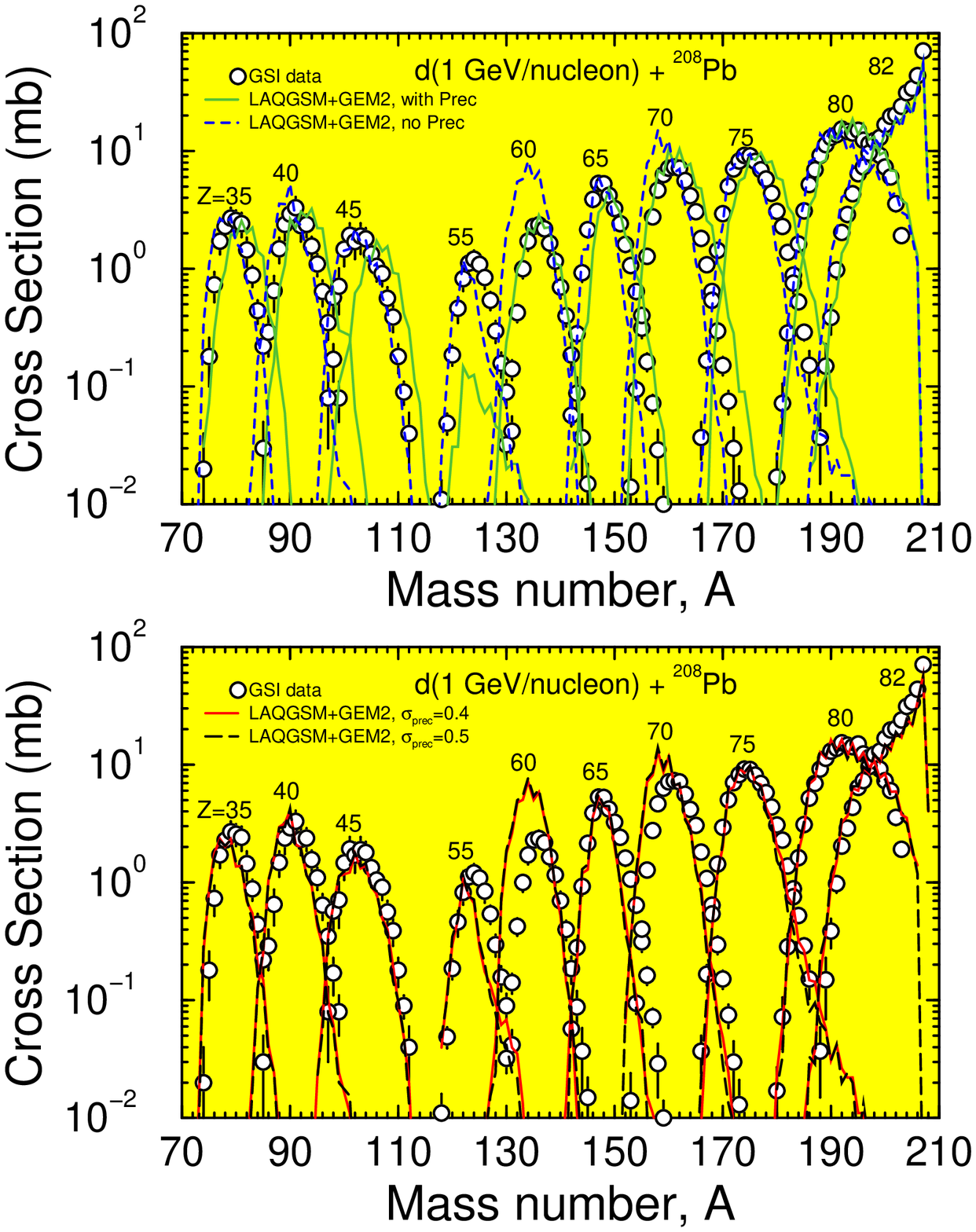}} 

\vspace{-5.5cm}
{\bf Figure 11.}
Experimental \cite{Enqvist02} mass distributions of the cross sections 
of ten isotopes with the charge $Z$ from 35 to 82
compared with our LAQGSM+GEM2 calculations. ``With Prec'' and ``no Prec'
results are shown on the upper plot, while results with reduced 
preequilibrium emission according to Eq. (9) using 
$\sigma_{prec} = 0.4$ and 0.5 are shown on the lower one.
\end{figure}

%\newpage
\begin{figure}

\vspace{-3.5cm}
\hspace{-25mm}
\includegraphics[width=12.5cm,angle=-90]{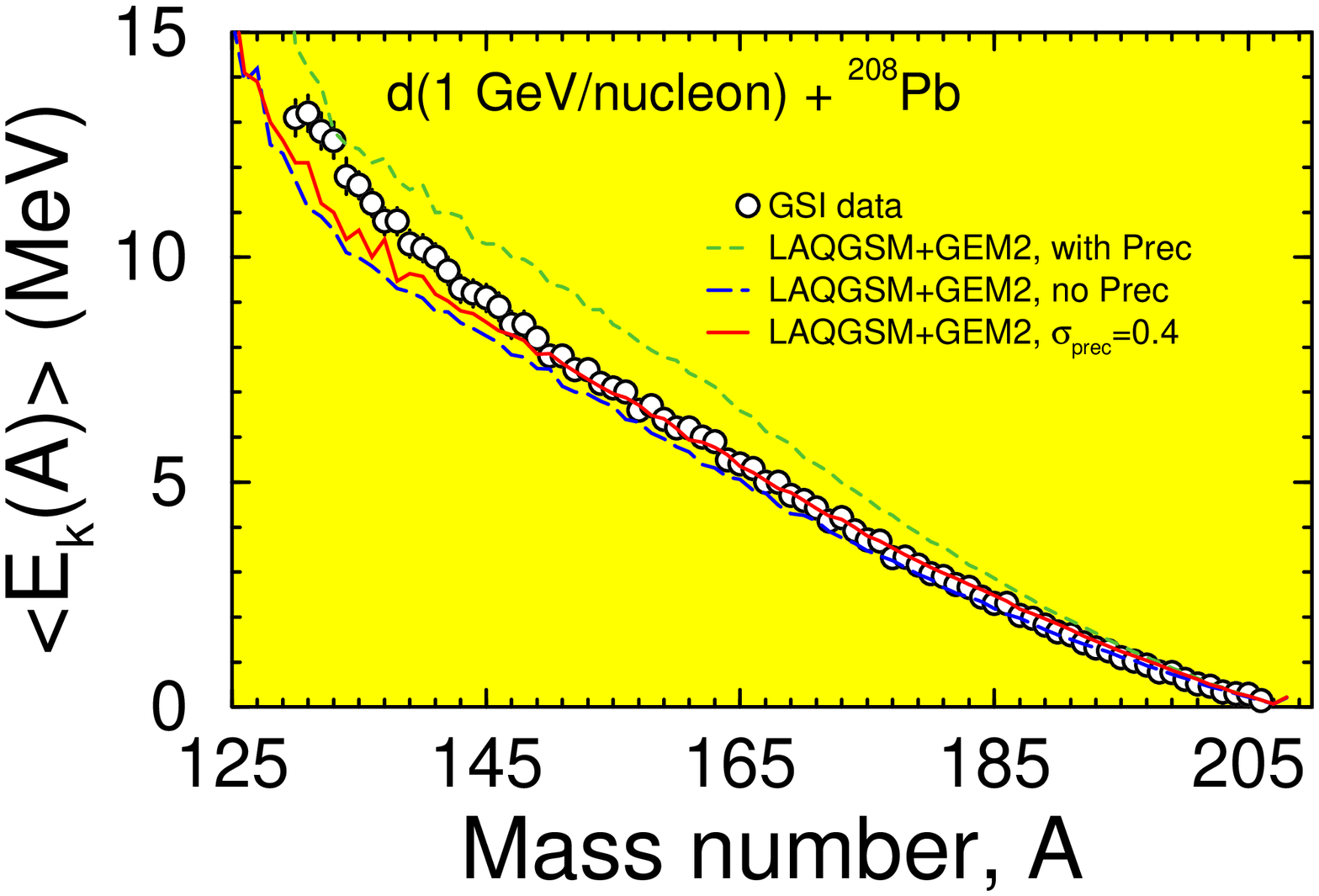}
{\bf Figure 12.}
Comparison of the experimental \cite{Enqvist02}
mass distribution of the
spallation-residue
kinetic energies of the 
nuclides produced in the reaction d(1 GeV/nucleon) + Pb (circles) with
the same LAQGSM+GEM2 calculations as shown in Fig. 10 .
\end{figure}

\newpage
\begin{figure}

\vspace{-10.0cm}
\centerline{\hspace{-0.7cm} \epsfxsize 23cm \epsffile{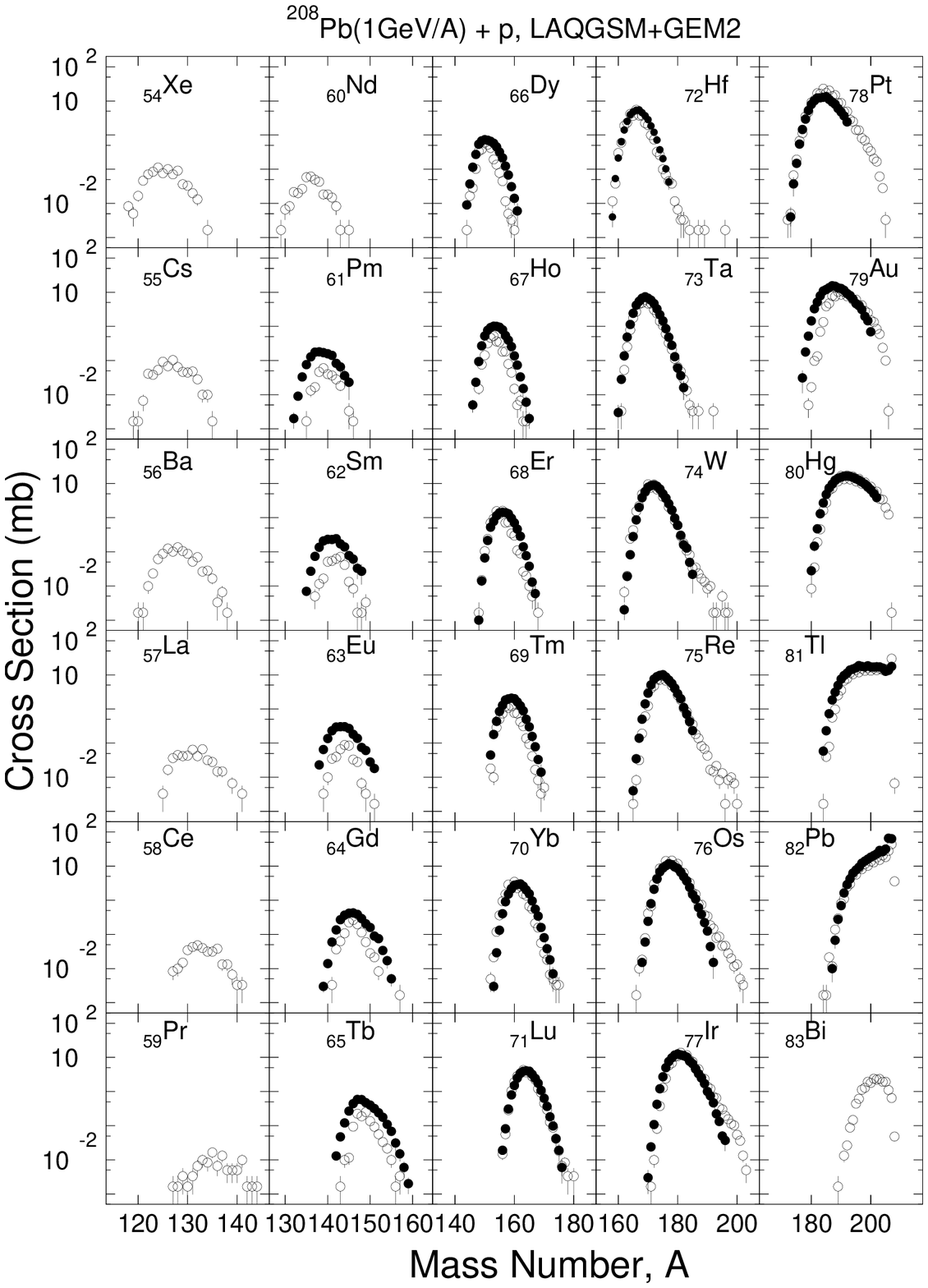}} 
\vspace{-2.cm}
{\bf Figure 13.}
Comparison of all measured \cite{Enqvist01} 
cross sections of spallation products from the reaction
1 GeV/nucleon $^{208}$Pb on p (black circles)
with our LAQGSM+GEM2 results (open circles).
\end{figure}

%\newpage
\begin{figure}

\vspace{-10.0cm}
\centerline{\hspace{-0.7cm} \epsfxsize 23cm \epsffile{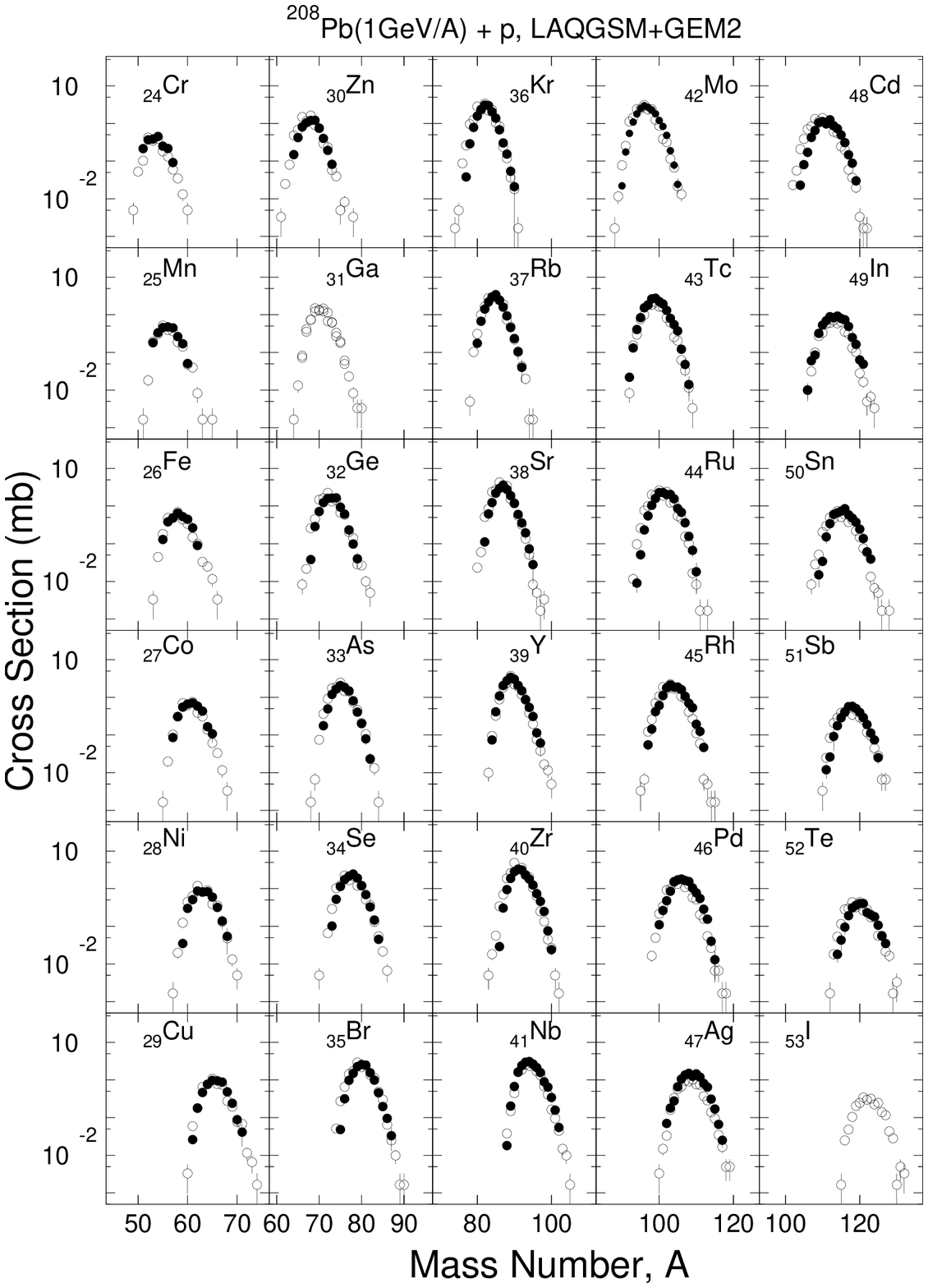}} 
\vspace{-2.cm}
{\bf Figure 14.}
Comparison of all measured \cite{Enqvist01} 
cross sections of fission products from the reaction
1 GeV/nucleon $^{208}$Pb on p (black circles)
with our LAQGSM+GEM2 results (open circles).

\end{figure}

%\newpage
\begin{figure}

\vspace{-10.0cm}
\centerline{\hspace{-0.7cm} \epsfxsize 23cm \epsffile{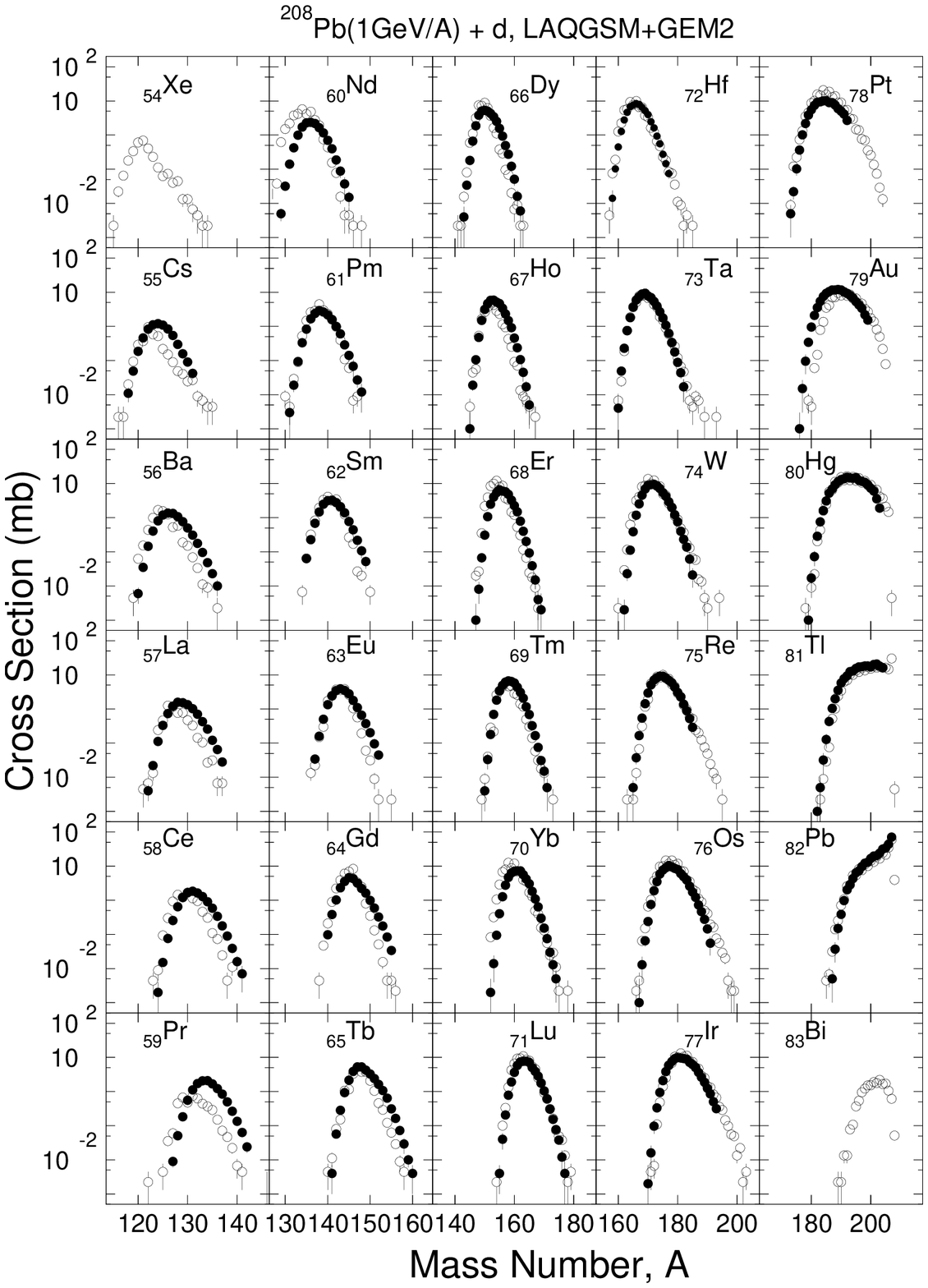}} 
%\centerline{\hspace{-0.7cm} \epsfxsize 20cm \epsffile{ubw.eps}} 
\vspace{-2.cm}
{\bf Figure 15.}
Comparison of all measured \cite{Enqvist02} 
cross sections of spallation products from the reaction
1 GeV/nucleon $^{208}$Pb on d (black circles)
with our LAQGSM+GEM2 results (open circles).
\end{figure}

%\newpage
\begin{figure}

\vspace{-10.0cm}
\centerline{\hspace{-0.7cm} \epsfxsize 23cm \epsffile{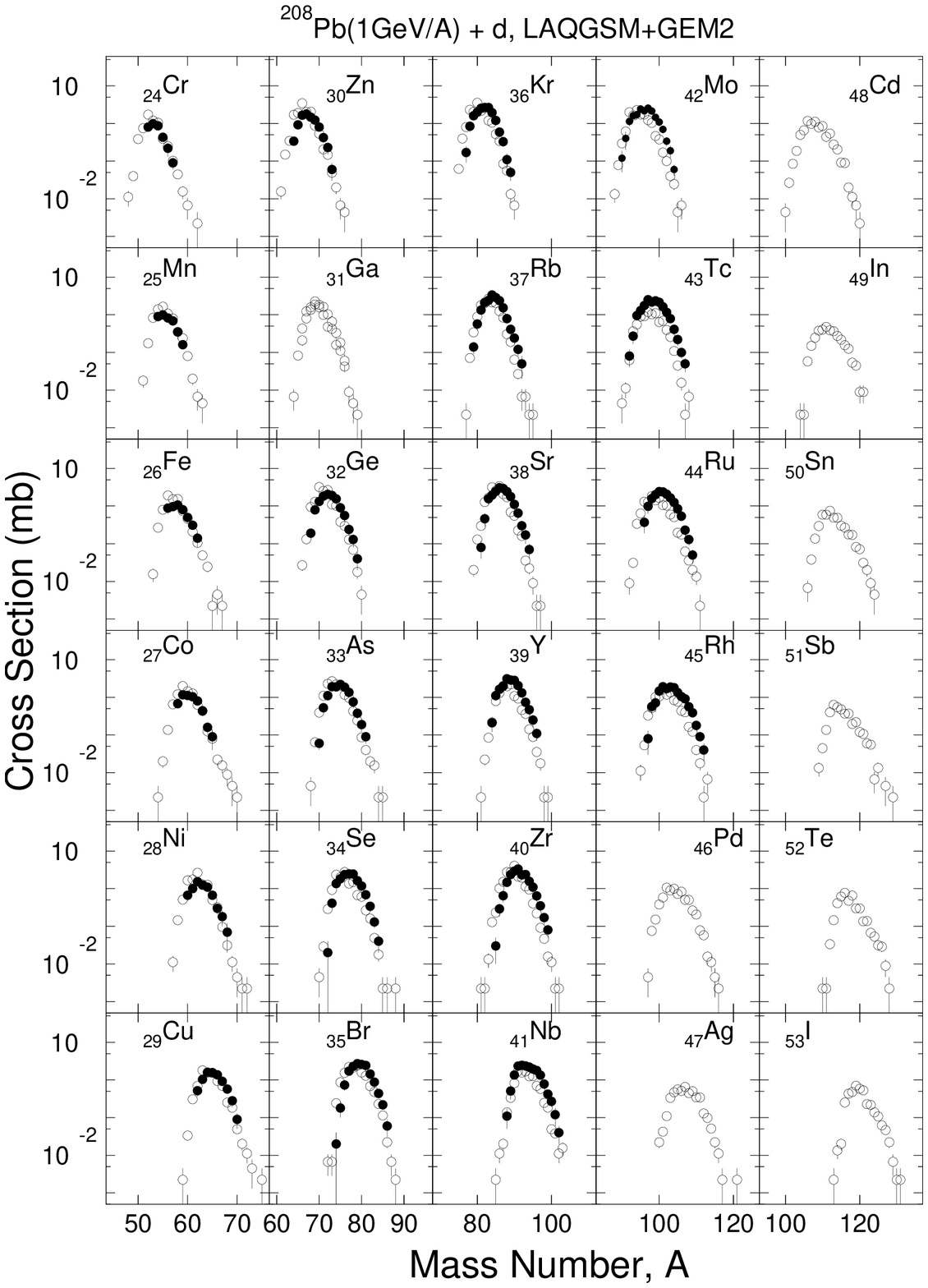}} 
%\centerline{\hspace{-0.7cm} \epsfxsize 20cm \epsffile{ubw.eps}} 
\vspace{-2.cm}
{\bf Figure 16.}
Comparison of all measured \cite{Enqvist02} 
cross sections of fission products from the reaction
1 GeV/nucleon $^{208}$Pb on d (black circles)
with our LAQGSM+GEM2 results (open circles).
\end{figure}

There is one more drawback of this approach to be mentioned: considering
evaporation of up to 66 particles in GEM becomes extremely time consuming 
when calculating reactions with heavy targets at high incident energies. 
But even this disadvantage may be mitigated by the performance of modern 
computers.
We have nevertheless some more serious doubts about the current version
of GEM2 related to its lack of self-consistency, 
{\it e.g.}:

1) using different, not physically related parameterizations for inverse
cross sections and Coulomb barriers for different particles and fragments;

2) using different level density parameters for the same compound nuclei
when calculating evaporation and estimating fission probability from the
widths of neutron evaporation and fission;

3) different, and purely phenomenological treatments of
fission for pre-actinide and actinide nuclei;

4) not taking into account at all the angular momentum of compound and
fissioning nuclei;

5) rough estimations for the fission barriers and level density parameters, 
{\it etc.}

This means that an approach like GEM2 can in principle be used to describe
fission and evaporation of particles and fragments heavier that $^4$He
after the INC and preequilibrium parts of CEM2k, LAQGSM, and other models,
but it should be considerably improved striving first to progressively
incorporate better physics, and only after that adjusting to selected data.

The results of the present work and from \cite{SantaFe02} show that
on the whole, the merged CEM2k+GEM2 and LAQGSM+GEM2 codes agree better 
with most of tested experimental data when we take 
into account the preequilibrium emission of particles, than 
when we neglect completely preequilibrium processes.
But there is still a not so clear question here; we have had some
indications for many years that CEM accounts for too many preequilibrium
particles, at least at energies above the pion-production
threshold. To solve this problem, as a ``zero-step'' approximation, in
our original CEM2k version \cite{CEM2k} we neglected in the exciton model
the transitions that decrease or do not change the number
of excitons $\Delta_n = -2$ and $\Delta_n = 0$, shortening in this way the
preequilibrium stage of reactions. We do not like this
approach as it is somehow arbitrary, ``ad hoc'', even though this
``never come back'' approximation is used in some popular codes
like LAHET \cite{LAHET} and FLUKA \cite{FLUKA}.
In the present work we removed this arbitrary
condition in CEM2k and the ``with Prec'' version takes into account all
the preequilibrium transitions $\Delta_n = +2$, 0, and -2, making
the preequilibrium stage of a reaction longer and increasing the
number of emitted preequilibrium particles. The results of the present
work indicate to us once again that we need to take into account the
preequilibrium stage in reactions, but we need less 
particle emission than we calculate at this stage with the standard
version of CEM.
In the present paper we address this problem reducing the number
of emitted preequilibrium particles following Veselsky \cite{sigpre},
as described above.

Besides GEM2, we have investigated the well known
code GEMINI by Charity \cite{GEMINI} as an alternative way to describe
production of various fragments by merging GEMINI with both our
CEM2k and LAQGSM, and we tested also the
thermodynamical fission model by Stepanov \cite{Stepanov}
with its own parameterizations for mass and charge widths, 
level-density parameters, fission barriers, {\it etc.}, merging it
with both CEM2k and LAQGSM to describe fission.
In addition, we have started to extend CEM2k and LAQGSM and to
develop our own fission model, as briefly noted in \cite{SantaFe02}.
The preliminary results we found for spallation, fission, and 
fragmentation products from several reactions we tested so far
using these approaches 
are very promising and we will present our results from these
studies in several separate papers.

\begin{center}
%\vspace{0.8cm}
{\it Acknowledgment}
\end{center}

\noindent

We thank Dr.\ Shiori Furihata for several useful discussions, for 
sending us her
Generalized Evaporation Model code GEM, and providing us further
with its updated version, GEM2.
We are grateful to  Dr. William B. Wilson
for providing us with results of his calculation the reaction 
p(100 MeV) + $^{238}$U with the codes CYF and YIELDX2k.
We thank Prof.\ Takashi Nakamura and Drs.\ Yoshiyuki Iwata and Hiroshi Iwase 
for sending us
numerical values of their measured neutron
spectra from heavy-ion reactions
and results of their calculations with QMD and HIC.

This study was supported by the U.\ S.\ Department of Energy and by the
Moldovan-U.\ S.\ Bilateral Grants Program, CRDF Project MP2-3025.

%\newpage

\end{document}